\documentclass[12pt,english,eqsecnum,prd,aps,nofootinbib,superscriptaddress,amsfonts,longbibliography,tightenlines]{revtex4-2}
\usepackage[T1]{fontenc}
\usepackage[utf8]{inputenc}
\usepackage{babel}
\usepackage{color}
\usepackage{mathrsfs}
\usepackage{mathtools}
\usepackage{amsmath}
\usepackage{amssymb}
\usepackage{stmaryrd}
\usepackage{graphicx}
\usepackage[unicode=true,pdfusetitle, bookmarks=true,bookmarksnumbered=false,bookmarksopen=false, breaklinks=true,pdfborder={0 0 1},backref=false,colorlinks=false] {hyperref}
\usepackage[normalem]{ulem}

\makeatletter
\allowdisplaybreaks[1]

\makeatother

\begin{document}

\title{Effective dynamics of the Schwarzschild  black hole interior with inverse triad corrections }

\author{Hugo A. Morales-T\'{e}cotl}
\email{hugo@xanum.uam.mx}
\affiliation{Departamento de F\'{i}sica, Universidad Aut\'{o}noma Metropolitana - Iztapalapa\\ San Rafael Atlixco 186, Ciudad de Mexico 09340, Mexico}
\affiliation{Departamento de F\'\i sica, Escuela Superior de F\'\i sica y Matem\'aticas del Instituto Polit\'ecnico Nacional \\
Unidad Adolfo L\'opez Mateos, Edificio 9, 07738 Ciudad de M\'exico, Mexico
}
\author{Saeed Rastgoo}
\email{srastgoo@yorku.ca}
\affiliation{School of Sciences and Engineering\\ Monterrey Institute of Technology (ITESM), Campus Le\'{o}n\\ Av. Eugenio Garza Sada, Le\'{o}n, Guanajuato 37190, Mexico}
\affiliation{Departamento de F\'{i}sica, Universidad Aut\'{o}noma Metropolitana - Iztapalapa\\ San Rafael Atlixco 186, Ciudad de Mexico 09340, Mexico}
\affiliation{Department of Physics and Astronomy, York University 4700 Keele Street, Toronto, Ontario M3J 1P3, Canada}
\author{Juan C. Ruelas}
\email{carlos.ruelas@tec.mx}
\affiliation{Departamento de F\'{i}sica, Universidad Aut\'{o}noma Metropolitana - Iztapalapa\\ San Rafael Atlixco 186, Ciudad de Mexico 09340, Mexico} 

\date{\today}
\begin{abstract}
We reconsider the study of the interior of the Schwarzschild black hole now including inverse triad quantum corrections within loop quantization. 
We derive these corrections and show that they are are related to  two parameters $\delta_b, \delta_c$   associated to the minimum length in the radial and angular directions, that enter Thiemann's trick for quantum  inverse triads.   Introduction of such corrections may lead to non-invariance of physical results under rescaling of the fiducial volume needed to compute the dynamics, due to noncompact topology of the model. So, we put forward two prescriptions to resolve this issue. These prescriptions amount to interchange $\delta_b, \delta_c$ in classical computations in Thiemann's trick.  By implementing the inverse triad corrections we found, previous results such as singularity resolution and black-to-white hole bounce hold with different values for the minimum radius-at-bounce, and the mass of the white hole.

\end{abstract}

\maketitle

\section{Introduction}

As one of the most fascinating predictions of general relativity,
black holes have been the subject of much analysis and explorations.
Particularly their interior, and the singularity located there, has
been studied in classical, quantum and semiclassical regimes. The
mainstream hope is that the classical singularity
will be resolved and replaced by a quantum region. However, there are still many open issues to be answered
in a satisfactory way. Within loop quantum gravity (LQG) \cite{Thiemann:2007pyv,Rovelli:2004tv},
there have been numerous works about quantum black holes and their
singularity resolution in both mini- and midi-superspace models, to
mention a few \cite{Ashtekar:2005qt,Boehmer:2008fz,Modesto:2005zm,Corichi:2015xia,Cartin:2006yv,Gambini:2009vp,Rastgoo:2013isa,Corichi:2015vsa,Corichi:2016nkp,Gambini:2009ie,Gambini:2013hna,Cortez:2017alh,RAY:2020}.
One of the most studied models in this context is the Schwarzschild
black hole which interior corresponds to a Kantowski-Sachs model, a system with finite degrees of freedom, and hence a mini-superspace,
with a singularity at the heart of it \cite{Boehmer:2008fz,Doran:2006dq}.
One of the approaches to quantize this model, inspired by LQG, is
polymer quantization \cite{Ashtekar:2002sn,Tecotl:2015cya,Morales-Tecotl:2016dma,Morales-Tecotl:2016ijb},
a technique also used in loop quantum cosmology (LQC) \cite{Ashtekar:2011ni,Ashtekar:2003hd,Ashtekar:2006uz}.
In this quantization the classical canonical algebra is represented
in a way that is unitarily inequivalent to the usual Schr\"{o}dinger representation
even at the kinematical level. The root of this inequivalency is the
choice of topology and the form of the inner product of this representation,
which renders some of the operators discontinuous in their parameters,
resulting in the representation not being weakly continuous. On the
other hand, unitary equivalency of a representation to the Schr\"{o}dinger
one is guaranteed by the Stone-von Neumann theorem iff all of its
premises, including weak continuity of the representation, are satisfied,
and the polymer representation does not. This inequivalency
translates into new results that are different from the usual quantization
of the system, one of them being the resolution of the singularity
of the Kantowski-Sachs model. These results however, are accompanied
by some issues that we briefly discuss in what follows. 

In one of the earliest attempts in this approach \cite{Ashtekar:2005qt},
the authors showed that the singularity can be avoided in the quantum
regime, but one of the important issues was the dependence of results
on auxiliary parameters that define the size of the fiducial cell.
The introduction of this fiducial cell, in this case a cylindrical
one with topology $\mathcal{I}\times\mathbb{S}^{2}$ and volume $V_{0}=a_{0}L_{0}$,
where $a_{0}$ is the area of the 2-sphere $\mathbb{S}^{2}$ and $L_{0}$
is the cylinder's height, is necessary to avoid the divergence of
some of the spatial integrals in homogenous models with some non-compact
directions. Particularly it is important to be able to define the
symplectic structure. Given that the physical results should not depend
on these auxiliary parameters, a new proposal, motivated by the ``improved
quantization'' in LQC \cite{Ashtekar:2006wn}, was put forward that
avoided this dependence and yielded bounded expansion and shear scalars
\cite{Boehmer:2008fz}. However, this method also leads to some
undesired modified behavior at the horizon due to quantum gravitational
effects in vacuum, that are manifestation of the coordinate singularity
there. There are also some other recent works that take a bit of a different approach to the problem by looking for an effective metric \cite{BenAchour:2018khr, Bojowald:2018xxu}.

In \cite{Corichi:2015xia}, a key modification to the quantization was proposed by choosing to
fix $a_{0}\coloneqq4\pi r_{0}^{2}$ by a physical scale $r_{0}$.
This physical $r_{0}$ permits one to define a Hamiltonian formulation,
and in this way is different in nature from the auxiliary scale $L_{0}$,
which is needed to fix the fiducial cell size to be able to define
the symplectic structure. Thus while $r_{0}$ will be present in physical
results in both the classical and quantum theories, these theories should
be independent of $L_{0}$. The proposal in \cite{Corichi:2015xia}, leads to results that
are independent of the auxiliary parameters, and while the theory
predicts that the singularity is resolved in the quantum gravity regime,
no large quantum gravitational effects appear at low curvatures near
the horizon as it should be the case.

It is worth noting that the anisotropic models suffered
from an issue: since these models resolve the singularity, they predicted
a ``bounce'' from a black hole to a white hole, with the mass $M_{W}$ 
of the resultant white hole not matching with that of the original
black hole $M_{B}\neq M_W$ , but rather $M_{W}\propto M_{B}^{4}$.
Recent work presented some proposals to deal with it \cite{Olmedo:2017lvt} and a different approach was developed in  \cite{Ashtekar:2018lag,Ashtekar:2018cay} by encompassing the interior region containing the classical singularity with the exterior asymptotic one, which, in the large mass limit, makes the masses of the white and black holes take the same value. See also \cite{Assanioussi:2019twp,Bodendorfer:2019jay,
Aruga:2019dwq} for a different perspective.

All previous works on black holes ignore  inverse triad corrections, to simplify the problem. However, they  are important especially at highly quantum
regimes and a more complete quantum gravitational analysis of this
model should take them into account. In this work  we first use a path integral
in phase space including inverse triad quantum corrections.  These corrections are known to produce severe
issues in non compact cosmological models, among them the dependence of the physical quantities on the
auxiliary parameters or their rescaling. We put forward two proposals that, in the case of Schwarzschild black hole, yield a physical description with no reference to fiducial parameters. Finally, we study how
these proposals modify the ``minimum radius at the bounce''. 

The structure of this paper is as follows: In section \ref{sec:Backg-classic},
we present the background, the relation between the Schwarzschild
interior and the Kantowski-Sachs model, and their classical Hamiltonian analysis. In section \ref{sec:quantum-H}, we briefly
review how the quantum Hamiltonian constraint is defined. In section
\ref{sec:PI-effective}, the path integral analysis is presented and
it is shown how a systematic effective Hamiltonian constraint can
be derived from the quantum Hamiltonian, including inverse triad corrections. Section \ref{sec:Issues-new-correc}
is dedicated to presenting some of the important issues that are raised
by the presence of the new corrections, and recognizing the root of
these issues. In section \ref{sec:Prescriptions}, we present two
proposals to deal with the aforementioned issues, and also show
their effect upon some physical quantities.
Finally, in section \ref{sec:Conclusion}, we conclude the paper by
presenting a summary and a discussion about the results. 

\section{Background and the classical theory\label{sec:Backg-classic}}

For Schwarzschild black hole the spacetime 
metric
\begin{equation}
ds^{2}=-\left(1-\frac{2GM}{r}\right)dt^{2}+\left(1-\frac{2GM}{r}\right)^{-1}dr^{2}+r^{2}\left(d\theta^{2}+\sin^{2}\theta d\phi^{2}\right)
\end{equation}
where $M$ is the mass of the black hole, the timelike
and spacelike curves switch their causal nature into each other for
observers that cross the event horizon. Hence the metric of the interior region is obtained by $r\leftrightarrow t$,
\begin{equation}
ds^{2}=-\left(\frac{2GM}{t}-1\right)^{-1}dt^{2}+\left(\frac{2GM}{t}-1\right)dr^{2}+t^{2}\left(d\theta^{2}+\sin^{2}\theta d\phi^{2}\right),\label{eq:sch-inter}
\end{equation}
with $t\in(0,2GM)$ and $r\in(-\infty,\infty)$. This metric is a
special case of a Kantowski-Sachs cosmological spacetime that is given
by the metric
\begin{equation}
ds^{2}=-d\tau^{2}+A^{2}(\tau)dr^{2}+B^{2}(\tau)\left(d\theta^{2}+\sin^{2}\theta d\phi^{2}\right).\label{eq:K-S-gener}
\end{equation}
The coordinates in which (\ref{eq:K-S-gener}) is written are 
Gaussian normal coordinates adapted
to the comoving observers, {\em i.e.}, the time coordinate curves are the
worldlines of the free falling objects ({\em e.g.} stars) that are at
rest with respect to such observers, and are parametrized
by their proper time $\tau$. The metric (\ref{eq:sch-inter}) can be seen to be derived
from (\ref{eq:K-S-gener}) by the transformation
\begin{equation}
d\tau^{2}=\left(\frac{2GM}{t}-1\right)^{-1}dt^{2}.
\end{equation}
Choosing positive root of the above, we get
\begin{equation}
\tau=-\sqrt{t(2GM-t)}-GM\tan^{-1}\left(\frac{t-GM}{\sqrt{t(2GM-t)}}\right)+\frac{GM\pi}{2},\label{eq:tau-in-terms-t-1}
\end{equation}
where the last term in the right hand side is the integration constant
and it is chosen such that $\tau\to0$ for $t\to0$ (at singularity),
and $\tau\to GM\pi$ for $t\to2GM$ (at the horizon), hence $\tau\in\left(0,GM\pi\right)$.
Then $\tau$ is a monotonic function of $t$. Written in Gaussian
normal coordinates $\left(\tau,r,\theta,\phi\right)$, the Schwarzschild
metric takes the form (\ref{eq:K-S-gener}), with $A^{2}(\tau)=\frac{2GM}{t(\tau)}-1$
and $B^{2}(\tau)=\tau^{2}$.

The Kantowski-Sachs metric (\ref{eq:K-S-gener}) in general and the
Schwarzschild interior (\ref{eq:sch-inter}) in particular, represent
a spacetime with spatial homogeneous but anisotropic foliations;  one
can consider $A(\tau)$ and $B(\tau)$ as two distinct scale
factors that affect the radial and angular parts of the metric separately.
Thus the interior region is a model with no local degrees of freedom, {\em i.e.}, it can be described as a mechanical
system with a finite number of configuration variables.
In gravitational language, this corresponds to a mini-superspace model.
From the computational point of view, this is
an important property that is exploited in quantizing the Schwarzschild
interior as we will see. 

It also can be seen from the metric (\ref{eq:K-S-gener}) and (\ref{eq:sch-inter}),
that the spacetime is foliated such that the spatial hypersurfaces have topology $\mathbb{R}\times\mathbb{S}^{2}$, and the symmetry
group is the Kantowski-Sachs isometry group $\mathbb{R}\times SO(3)$.
The aforementioned topology of the model means that there exists one
noncompact direction, $r\in\mathbb{R}$ in space. Thus in order to
be able to compute quantities that involve integrals over space, particularly
the symplectic structure $\int_{\mathbb{R}\times\mathbb{S}^{2}}\text{d}^{3}x\,\text{d}q\wedge\text{d}p$,
one needs to choose a finite fiducial volume over which these integrals
are calculated, otherwise the integrals will diverge. This is a common
practice in the study of homogeneous minisuperspace models, which
here, is done by introducing an auxiliary length $L_{0}$ to restrict
the noncompact direction to an interval $r\in\mathcal{I}=[0,L_{0}]$.
The volume of the fiducial cylindrical cell in this case is $V_{0}=a_{0}L_{0}$,
where $a_{0}$ is the area of the 2-sphere $\mathbb{S}^{2}$ in $\mathcal{I}\times\mathbb{S}^{2}$. Now, for the area $a_{0}$,
there are at least two choices: One can use it as an auxiliary parameter,
or fix it using a physical scale. In any case, the final physical
results should not depend on the choice of auxiliary parameters. In
a recent work \cite{Corichi:2015xia}, a choice has been put forward,
in which the $\mathbb{S}^{2}$ area of the fiducial volume is fixed
to be $a_{0}\coloneqq4\pi r_{0}^{2}$ where $r_{0}$ is a physical
scale that is identified with the Schwarzschild radius. This choice
can be considered as a boundary condition which ensures that the classical
limit becomes the classical Schwarzschild solution with radius $r_{0}$.
Using this choice, the volume of the cylindrical fiducial cell becomes
$V_{0}=4\pi r_{0}^{2}L_{0}$, and the associated fiducial metric is
denoted by $^{0}q_{ab}$. Using a physical scale for $a_{0}$, instead
of an auxiliary nonphysical one, seems to be a key ingredient that
fixes some of the issues with previous attempts at loop quantization of
the interior of Schwarzschild black hole, and here we follow this choice. 

The starting point of the Hamiltonian analysis in this approach is
to write down the classical configuration variable, the $su(2)$ Ashtekar-Barbero
connection $A_{a}^{i}$, and its conjugate momentum, the desitized triad $E_{i}^{a}$,
in the relevant coordinate basis. Given the symmetries of this spacetime
and after imposing the Gauss constraint, these variables take the
form \cite{Ashtekar:2005qt,Corichi:2015xia} 
\begin{align}
A_{a}^{i}\tau_{i}dx^{a}= & \bar{c}\tau_{3}dr+\bar{b}r_{0}\tau_{2}d\theta-\bar{b}r_{0}\tau_{1}\sin\theta d\phi+\tau_{3}\cos\theta d\phi,\label{eq:A-AB}\\
E_{i}^{a}\tau^{i}\frac{\partial}{\partial x^{a}}= & \bar{p}_{c}r_{0}^{2}\tau_{3}\sin\theta\frac{\partial}{\partial r}+\bar{p}_{b}r_{0}\tau_{2}\sin\theta\frac{\partial}{\partial\theta}-\bar{p}_{b}r_{0}\tau_{1}\frac{\partial}{\partial\phi},\label{eq:E-AB}
\end{align}
where $\bar{b}$, $\bar{c}$, $\bar{p}_{b}$ and $\bar{p}_{c}$ are
functions that only depend on time $t$, and $\tau_{i}=-i\sigma_{i}/2$
are a $su(2)$ basis with $\sigma_{i}$ being the Pauli matrices.
 $r_{0}=2GM$ is the Schwarzschild radius.
In these variables the Schwarzschild interior metric becomes
\begin{equation}
ds^{2}=-N^{2}dt^{2}+\frac{\bar{p}_{b}^{2}}{\bar{p}_{c}}dr^{2}+\bar{p}_{c} r_{0}^{2}(d\theta^{2}+\sin^{2}\theta d\phi^{2}),
\end{equation}
where we have chosen $\bar{p}_{c}\geq 0$ given that it is related to the radial coordinate (see below). Now, the symplectic structure can be computed by performing an integration
over the fiducial volume as
\begin{align}
\Xi= & \frac{1}{8\pi G\gamma}\int_{\mathcal{I}\times\mathbb{S}^{2}}\text{d}^{3}x\quad\text{d}A_{a}^{i}\wedge\text{d}E_{i}^{a}\nonumber \\
= & \frac{L_{0}r_{0}^{2}}{2G\gamma}\left(\text{d}\bar{c}\wedge\text{d}\bar{p}_{c}+2\text{d}\bar{b}\wedge\text{d}\bar{p}_{b}\right),
\end{align}
where $\gamma$ is the Barbero-Immirzi parameter \cite{Thiemann:2007pyv}.
Clearly, in these variables, the symplectic structure and thus the
Poisson algebra depends on $L_{0}$. To remove this dependency, it
is customary to redefine the variables in the following way
\begin{equation}
c=L_{0}\bar{c},\quad p_{c}=r_{0}^{2}\bar{p}_{c}\quad b=r_{0}\bar{b}\quad p_{b}=r_{0}L_{0}\bar{p}_{b}.\label{eq:redef-b-c-pb-pc}
\end{equation}
As a result the Poisson algebra  between these redefined variables, and other \emph{physical} quantities
are explicitly independent of the auxiliary variable $L_{0}$, 
\begin{equation}
\{c,p_{c}\}=2G\gamma,\quad\quad\{b,p_{b}\}=G\gamma,
\end{equation}
and the physical metric takes the form
\begin{equation}
ds^{2}=-N^{2}dt^{2}+\frac{p_{b}^{2}}{L_{0}^{2} p_{c}}dx^{2}+ p_{c}(d\theta^{2}+\sin^{2}\theta d\phi^{2}).
\end{equation}
By comparing this metric with (\ref{eq:sch-inter}), and assuming
we are working in Schwarzschild coordinates, we can see that 
\begin{equation}
\frac{p_{b}^{2}}{L_{0}^{2} p_{c}}=\left(\frac{2GM}{t}-1\right),\,\,\,\,\,\,\,\,\,\,\,\,\,\,\,\,\,\,\vert p_{c}\vert=t^{2}.\label{eq:Sch-corresp}
\end{equation}
This means that
\begin{align}
p_{b}= & 0, & p_{c}= & 4G^{2}M^{2}, &  & \textrm{On the horizon\,}t=2GM\,(\tau=GM\pi),\label{eq:t-horiz}\\
p_{b}\to & 0, & p_{c}\to & 0, &  & \textrm{At singularity\,}t=0 \, (\tau=0),\label{eq:t-singular}
\end{align}
where $t$ is the time in Schwarzschild coordinates, and we have used
the Schwarzschild lapse $N=\left(\frac{2GM}{t}-1\right)^{-\frac{1}{2}}$
to find the corresponding proper times $\tau=\int Ndt\in\left(0,GM\pi\right)$. 

Although the redefinitions (\ref{eq:redef-b-c-pb-pc}), transform
the metric such that it remains invariant under coordinate rescaling
$r\to\xi r$, there still exists a freedom in rescaling the length
of the interval $\mathcal{I}$ itself by $L_{0}\rightarrow\xi L_{0}$.
This freedom manifests itself in transformation of the canonical variables
in the following way 
\begin{eqnarray}
c\rightarrow c^{\prime}=\xi c & \quad\quad & p_{c}\rightarrow p_{c}^{\prime}=p_{c},\label{eq:resc-c-pc}\\
b\rightarrow b^{\prime}=b & \quad\quad & p_{b}\rightarrow p_{b}^{\prime}=\xi p_{b}.\label{eq:resc-b-pb}
\end{eqnarray}
Note that here \cite{Corichi:2015xia}, since $r_{0}$
is chosen to be a physical scale, not an auxiliary one, there is no
freedom associated with its rescaling, unlike the case in \cite{Ashtekar:2005qt}.

\section{The quantum Hamiltonian constraint\label{sec:quantum-H}}

The next step is to find the classical Hamiltonian in loop variables,
and then representing it as an operator on a suitable kinematical
Hilbert space. We only briefly go over this,  details can
be found in previous works \cite{Ashtekar:2005qt,Cartin:2006yv}.
Since in this model, the diffeomorphism constraint is trivially satisfied,
after imposing the Gauss constraint, one is left only with the classical
Hamiltonian constraint
\begin{equation}
C=-\int\text{d}^{3}x\frac{N}{\sqrt{\vert\text{det}E\vert}}\epsilon_{ijk}E^{ai}E^{bj}\left(\frac{1}{\gamma^{2}}{}^{0}F_{ab}^{k}-\Omega_{ab}^{k}\right).\label{eq:H-constr}
\end{equation}
Here the integral is over the fiducial volume, and $\Omega_{ab}^{k}$
and $^{0}F_{ab}^{k}$ are the curvatures of the spin connection $\Gamma_{a}^{i}$,
and the extrinsic curvature $K_{a}^{i}=\frac{1}{\gamma}\left(A_{a}^{i}-\Gamma_{a}^{i}\right)$,
respectively. Since
in loop quantum gravity the configuration variables are holonomies,
not the connections themselves, these curvatures should be written
in terms of them. In general, the holonomy of a connection
$A_{a}^{i}$ over edge $e$ is the path ordered exponential
\begin{equation}
h_{e}[A]=\mathcal{P}\textrm{exp}\left(\int_{e}A_{a}^{i}\tau_{i}dx^{a}\right).
\end{equation}
In case of the present model, there are two types of holonomies: the
one that is integrated over a path (or edge) $\lambda$, in the $r$
direction,
\begin{equation}
h_{r}^{(\lambda)}=\cos\left(\frac{\lambda c}{2}\right)+2\tau_{3}\sin\left(\frac{\lambda c}{2}\right)\label{eq:holon-x}
\end{equation}
and the ones that are over edges $\mu$, in $\theta$ and $\phi$
directions,
\begin{align}
h_{\theta}^{(\mu)}= & \cos\left(\frac{\mu b}{2}\right)+2\tau_{2}\sin\left(\frac{\mu b}{2}\right),\label{eq:holon-theta}\\
h_{\phi}^{(\mu)}= & \cos\left(\frac{\mu b}{2}\right)-2\tau_{1}\sin\left(\frac{\mu b}{2}\right).\label{eq:holon-phi}
\end{align}
To find the curvature, one considers loops in $r-\theta$, $r-\phi$
and $\theta-\phi$ planes, such that the edges along the $r$ direction
in $\mathbb{R}$ have a length $\delta_{c}\ell_{c}$ where $\ell_{c}=L_{0}$,
and the edges along the longitude and the equator of $\mathbb{S}^{2}$
have length $\delta_{b}\ell_{b}$ where $\ell_{b}=r_{0}$. These lengths $\delta_b\ell_b,\delta_c\ell_c$ are considered with respect to the fiducial metric
\begin{equation}
ds_{0}^{2}:=dr^{2}+r_{0}^{2}(d\theta^{2}+\sin^{2}\theta d\phi^{2}).
\end{equation} 
Then the
curvature $^{0}F_{ab}^{k}$ can be computed in terms of holonomies
as
\begin{equation}
^{0}F_{ab}^{k}=-2\lim_{\textrm{Ar}\boxempty\rightarrow0}\text{Tr}\left(\frac{h_{\boxempty_{ij}}^{(\delta_{(i)},\delta_{(j)})}-1}{\delta_{(i)}\ell_{(i)}\delta_{(j)}\ell_{(j)}}\tau^{k}\right){}^{0}E_{a}^{i}{}^{0}E_{b}^{j},\label{eq:class-curvatr}
\end{equation}
in which
\begin{equation}
h_{\boxempty_{ij}}^{(\delta_{(i)},\delta_{(j)})}=h_{i}^{(\delta_{(i)})}h_{j}^{(\delta_{(j)})}\left(h_{i}^{(\delta_{(i)})}\right)^{-1}\left(h_{j}^{(\delta_{(j)})}\right)^{-1}.
\end{equation}
Here $\delta_{(i)}$ correspond to $\delta_{b}$ or $\delta_{c}$,
$\ell_{(i)}$ correspond to $\ell_{(b)}$ or $\ell_{(c)}$, $\boxempty_{ij}$
is the loop with edges $i,j$, and $\textrm{Ar}\boxempty$ is the
area of the loop over which the curvature is being computed, and its
limit to zero has been taken. Note, however, that in the quantum regime,
due to the discreteness of the area, the loops can only be shrunk
to a minimum value of $\Delta=\zeta\ell_{\textrm{Pl}}^{2}$ with $\zeta\approx\mathcal{O}(1)$
\cite{Ashtekar:2005qt}. The choice in
\cite{Corichi:2015xia} is such that
\begin{align}
\delta_{b}= & \frac{\sqrt{\Delta}}{r_{0}}, & \delta_{c}= & \frac{\sqrt{\Delta}}{L_{0}}.\label{eq:delta-Delta}
\end{align}
As for the factor outside the parenthesis in (\ref{eq:H-constr}), which
contains the inverse triad, we rewrite it using Thiemann's trick
\begin{equation}
\frac{\epsilon_{ijk}}{\sqrt{\vert\text{det}E\vert}}E^{aj}E^{bk}=\sum_{k}\frac{^{0}\epsilon^{abc}\,{}^{0}E_{c}^{k}}{2\pi\gamma G\delta_{(k)}\ell_{(k)}}\mathrm{Tr}\left(h_{k}^{(\delta_{(k)})}\left\{ \left(h_{k}^{(\delta_{(k)})}\right)^{-1},V\right\} \tau_{i}\right),\label{eq:thiemann-trick}
\end{equation}
in which $V$, the physical volume of the fiducial cell, is
\begin{equation}
V=\int d^{3}x\sqrt{\det q}=4\pi\,|p_{b}||p_{c}|^{1/2}.\label{eq:class-vol}
\end{equation}
The reason for writing the left hand side of \ref{eq:thiemann-trick} in the rather complicated form of 
its right hand side is the problem with representing the complicated expression $\frac{1}{\sqrt{\vert\text{det}E\vert}}$, which contains a fraction as well as a square root. Hence one uses the Thiemann's trick by writing it as a Poisson bracket, that can then be turned into a bracket in quantization procedure using the Dirac prescription $\{A,B\}\to i\left[\hat{A},\hat{B}\right]$ where no complication will arise and the issue is bypassed. Another more technical and more important reason is that the (discrete) spectrum of the area operator in loop quantum gravity includes 0 and thus the inverse operator is not properly defined on eigenstates with vanishing area. In summary, Thiemann's trick helps us hide the non-polynamiality of the theory and obtain a finite Hamiltonian.

Considering that $\Omega_{ab}^{k}$ in this case becomes 
\begin{equation}
\Omega=-\sin\left(\theta\right)\tau^{3}d\theta\wedge d\phi
\end{equation}
and using (\ref{eq:class-curvatr}), (\ref{eq:thiemann-trick}) and (\ref{eq:class-vol})
in (\ref{eq:H-constr}), we get 
\begin{align}
C=NC^{\left(\delta_{b},\delta_{c}\right)}= & -\frac{2N}{\gamma^{3}G\delta_{b}^{2}\delta_{c}}\left[2\gamma^{2}\delta_{b}^{2}\mathrm{Tr}\left(\tau_{3}h_{x}^{(\delta_{c})}\left\{ \left(h_{x}^{(\delta_{c})}\right)^{-1},V\right\} \right)\right.\nonumber \\
 & \left.+\sum_{ijk}\epsilon^{ijk}\mathrm{Tr}\left(h_{\Box_{ij}}^{\left(\delta_{(i)},\delta_{(j)}\right)}h_{k}^{\left(\delta_{(k)}\right)}\left\{ \left(h_{k}^{(\delta_{(k)})}\right)^{-1},V\right\} \right)\right].\label{eq:H-constr-deltas}
\end{align}
To construct the kinematical Hilbert space on which this Hamiltonian
constraint is to be represented, one notes that the algebra generated
by the holonomies (\ref{eq:holon-x})-(\ref{eq:holon-phi}), is the
algebra of the almost periodic functions of the form $\exp(i(\mu b+\lambda c)/2)$.
This algebra (for just $b$ or $c$) is isomorphic to the algebra
of the continuous functions on the Bohr compactification of $\mathbb{R}$.
Thus the kinematical Hilbert space corresponding to this space of
configurations turns out to be the Cauchy completion of the space
of square integrable functions over the Bohr compactified $\mathbb{R}^{2}$,
together with its associated Haar measure $\mathscr{H}_{\textrm{kin}}=L^{2}(\mathbb{R}_{Bohr}^{2},d^2\mu_{Bohr})$.
The basis states of this space satisfy the relation
\begin{equation}
\langle\mu^{\prime},\lambda^{\prime}\vert\mu,\lambda\rangle=\delta_{\mu,\mu^{\prime}}\delta_{\lambda,\lambda^{\prime}},
\end{equation}
where on the right hand side we have Kronecker deltas instead of 
Dirac deltas. On this space, in the momentum basis, the basic variables
are represented as 
\begin{align}
 & \widehat{e^{\frac{1}{2}i\delta_{b}b}}\vert\mu,\lambda\rangle=\vert\mu+\delta_{b},\lambda\rangle, &  & \widehat{e^{\frac{1}{2}i\delta_{c}c}}\vert\mu,\lambda\rangle=\vert\mu,\lambda+\delta_{c}\rangle,\label{eq:exp-reps}\\
 & \hat{p}_{b}\vert\mu,\lambda\rangle=\frac{\gamma\ell_{\text{Pl}}^{2}}{2}\mu\vert\mu,\lambda\rangle, &  & \hat{p}_{c}\vert\mu,\lambda\rangle=\gamma\ell_{\text{Pl}}^{2}\lambda\vert\mu,\lambda\rangle.\label{eq:p-rep}
\end{align}
This is the quantum mechanical polymer representation corresponding to the original classical variables,
which is unitarily inequivalent to the Schr\"{o}dinger representation,
due to some of the operators not being weakly continuous in their
parameters, and hence the representation not satisfying the weak continuity
premise of the Stone-von Neumann theorem. Due to the lack of weak continuity,
the operators $\hat{b}$ and $\hat{c}$ are not well-defined on $\mathscr{H}_{\textrm{kin}}$,
and thus their corresponding infinitesimal transformations do not
exist. The theory, thus, only contains their corresponding finite
transformations due to the action of $\widehat{e^{\frac{1}{2}i\delta_{b}b}}$
and $\widehat{e^{\frac{1}{2}i\delta_{c}c}}$, which are not to be
considered as the literal exponentiation of $\hat{b}$ and $\hat{c}$.
This finite transformation is evident from (\ref{eq:exp-reps}). These
result in $p_{b}$ and $p_{c}$ (components of the triad $E_{i}^{a}$)
being discrete in the sense that they can only change by a finite
minimum value. This, in principle, is how this approach yields the
quantization and discreteness of the geometry. 

Using the above consideration, the Hamiltonian constraint (\ref{eq:H-constr-deltas}),
is represented as
\begin{align}
\hat{C}^{\left(\delta_{b},\delta_{c}\right)}= & \frac{32i}{\gamma^{3}\delta_{b}^{2}\delta_{c}\ell_{\textrm{Pl}}^{2}}\left\{ \left[\sin\left(\frac{\delta_{b}b}{2}\right)\cos\left(\frac{\delta_{b}b}{2}\right)\sin\left(\frac{\delta_{c}c}{2}\right)\cos\left(\frac{\delta_{c}c}{2}\right)\right]\right.\nonumber \\
 & \times\left[\sin\left(\frac{\delta_{b}b}{2}\right)\hat{V}\cos\left(\frac{\delta_{b}b}{2}\right)-\cos\left(\frac{\delta_{b}b}{2}\right)\hat{V}\sin\left(\frac{\delta_{b}b}{2}\right)\right]\nonumber \\
 & +\frac{1}{2}\left[\sin^{2}\left(\frac{\delta_{b}b}{2}\right)\cos^{2}\left(\frac{\delta_{b}b}{2}\right)+\frac{1}{4}\gamma^{2}\delta_{b}^{2}\right]\nonumber \\
 & \left.\times\left[\sin\left(\frac{\delta_{c}c}{2}\right)\hat{V}\cos\left(\frac{\delta_{c}c}{2}\right)-\cos\left(\frac{\delta_{c}c}{2}\right)\hat{V}\sin\left(\frac{\delta_{c}c}{2}\right)\right]\right\} ,
\end{align}
where $\hat{V}$ is the quantum volume operator, which is the representation
of the classical volume (\ref{eq:class-vol}), on $\mathscr{H}_{\textrm{kin}}$.
We will consider rather the symmetric version of the above operator $\hat{C}_{\textrm{S}}^{\left(\delta_{b},\delta_{c}\right)}=\frac{1}{2}\left(\hat{C}^{\left(\delta_{b},\delta_{c}\right)}+\hat{C}^{\left(\delta_{b},\delta_{c}\right)\dagger}\right)$.

\section{Path integral analysis: effective Hamiltonian and new features\label{sec:PI-effective}}

To find the effective version of the constraint we employ path integration. For standard mechanical systems path integrals yield expressions for the matrix elements of the evolution operators. The original derivation by Feynman involved the canonical theory expressing the evolution by composing $\mathcal{N}$ infinitesimal ones and inserting complete basis between these. Such discrete time path integral gets replaced by the continuum one in the limit $\mathcal{N}\rightarrow \infty$. For gravitational models we have two different, but equivalent, routes to follow \cite{Ashtekar:2010gz} (See \cite{Morales-Tecotl:2016ijb,Tecotl:2015cya} for the case of non gravitational models): the use of a relational or deparametrized time scheme in which a matter degree of freedom is used as a clock, or else consider a timeless scheme that may include matter. In the latter case there is no evolution operator but a constraint and its solutions, and we consider this scheme next. The aim is to construct a path integral expression for the so called ``extraction amplitude'' \cite{Ashtekar:2010gz}
\begin{equation}
A\left(\mu_{f},\lambda_{f};\mu_i, \lambda_{i}\right)=
\int d\alpha \left\langle \mu_{f},\lambda_{f}\left| \mathrm{e}^
{-i \alpha\hat{C}_{\textrm{S}}^{\left(\delta_{b},\delta_{c}\right)}}   \right|\mu_{i},\lambda_{i} \right\rangle ,
\end{equation}
which is a Green function for the transformation between kinematical and physical states, those that are annihilated by the quantum Hamiltonian constraint,
\begin{equation}
|\Psi_{\mathrm{phys}} \rangle = \int d\alpha \;
\mathrm{e}^
{-i\alpha\hat{C}_{\textrm{S}}^{\left(\delta_{b},\delta_{c}\right)}} 
|\Psi_{\mathrm{kin}} \rangle
\end{equation}
with $|\Psi_{\mathrm{kin}} \rangle \in \mathscr{H}_{\textrm{kin}}$, namely
\begin{equation}
\Psi_{\mathrm{phys}} (\mu,\lambda)
= \sum_{\lambda',\mu'} 
A\left(\mu,\lambda; \mu',\lambda' \right)
\Psi_{\mathrm{kin}} (\mu',\lambda').
\end{equation}
To find the path integral representation
of this Green function, as usual, we employ the ``time slicing'' method \cite{Ashtekar:2010gz}, by dividing
the fictitious unit time interval into $\mathcal{N}$ sub-intervals each with
length $\epsilon=\frac{1}{\mathcal{N}}$ such that $\mathrm{e}^
{-i \alpha\hat{C}_{\textrm{S}}^{\left(\delta_{b},\delta_{c}\right)}} = \mathrm{e}^
{-i t\hat{H}}, \; \hat{H}:= \alpha\hat{C}_{\textrm{S}}^{\left(\delta_{b},\delta_{c}\right)},\; t=1,$ and hence
\begin{equation}
A\left(\mu_{f},\lambda_{f},\mu,\lambda_{i} \right)=\int d\alpha\left\langle \mu_{f},\lambda_{f}\left|\underbrace{\mathrm{e}^{-i\epsilon \alpha 
\hat{C}_{\textrm{S}}^{\left(\delta_{b},\delta_{c}\right)} }
\ldots
\mathrm{e}^{-i\epsilon \alpha 
\hat{C}_{\textrm{S}}^{\left(\delta_{b},\delta_{c}\right)} }
}_{\mathcal{N}\,\textrm{times}}\right|\mu_{i},\lambda_{i}\right\rangle .
\end{equation}
By inserting $\hat{\mathbb{I}}=\sum_{(\mu,\lambda)\in\Gamma}\vert\mu,\lambda\rangle\langle\mu,\lambda\vert$,
between the exponentials above, the amplitude is written as
\begin{equation}
A\left(\mu_{f},\lambda_{f},\mu,\lambda_{i} \right)= \int d\alpha \prod_{n=1}^{\mathcal{N}}\sum_{\mu_{n},\lambda_{n}\in\gamma}\left\langle \mu_{n},\lambda_{n}\left|
\mathrm{e}^{-i\epsilon \alpha
\hat{C}_{\textrm{S}}^{\left(\delta_{b},\delta_{c}\right)} }
\right|
\mu_{n-1},\lambda_{n-1}\right\rangle ,\label{eq:Kernel-1}
\end{equation}
where $\mu_{\mathcal{N}},\lambda_{\mathcal{N}},\mu_{0},\lambda_{0}$
correspond to $\mu_{f},\lambda_{f},\mu_{i},\lambda_{i}$ respectively.
Each ``short-time'' amplitude can be expanded up to first order in $\epsilon$
as
\begin{align}
\left\langle \mu_{n},\lambda_{n}\left|e^{-i\epsilon\alpha \hat{C}_{\textrm{S}}^{\left(\delta_{b},\delta_{c}\right)}}\right|\mu_{n-1},\lambda_{n-1}\right\rangle = & \delta_{\mu_{n},\mu_{n-1}}\delta_{\lambda_{n},\lambda_{n-1}}-i\epsilon\left\langle \mu_{n},\lambda_{n}\left|\hat{C}_{\textrm{S}}^{\left(\delta_{b},\delta_{c}\right)}\right|\mu_{n-1},\lambda_{n-1}\right\rangle +\mathcal{O}(\epsilon^{2}),\nonumber \\
= & \left(\frac{1}{2\pi}\right)^{2}\int_{-\pi/p_{b}^{0}}^{\pi/p_{b}^{0}}db_{n}\int_{-\pi/p_{c}^{0}}^{\pi/p_{c}^{0}}dc_{n}\,e^{-ib_{n}\left(p_{n}^{b}-p_{n-1}^{b}\right)-ic_{n}\left(p_{n}^{c}-p_{n-1}^{c}\right)}\nonumber \\
 & -i\epsilon \alpha \left\langle \mu_{n},\lambda_{n}\left|\hat{C}_{\textrm{S}}^{\left(\delta_{b},\delta_{c}\right)}\right|\mu_{n-1},\lambda_{n-1}\right\rangle +\mathcal{O}(\epsilon^{2}),\label{eq:Exp-Pi-1}
\end{align}
where we have used $p_{n}^{b}=\frac{1}{2}\gamma\ell_{\text{Pl}}^{2}\mu_{n}$
and $p_{n}^{c}=\gamma\ell_{\text{Pl}}^{2}\lambda_{n}$.

To proceed, we need to compute the matrix element of the quantum Hamiltonian
constraint $\hat{C}_{\textrm{S}}^{\left(\delta_{b},\delta_{c}\right)}$.
This turns out to be
\begin{align}
\left\langle \mu^{\prime},\lambda^{\prime}\left|\hat{C}_{\textrm{S}}^{\left(\delta_{b},\delta_{c}\right)}\right|\mu,\lambda\right\rangle = & -\frac{1}{\gamma^{3}\delta_{b}^{2}\delta_{c}\ell_{\textrm{Pl}}^{2}}\left[\left(V_{\mu+\delta_{b},\lambda}-V_{\mu-\delta_{b},\lambda}\right)\right.\nonumber \\
 & \times\left(\delta_{\mu^{\prime},\mu+2\delta_{b}}-\delta_{\mu^{\prime},\mu-2\delta_{b}}\right)\left(\delta_{\lambda^{\prime},\lambda+2\delta_{c}}-\delta_{\lambda^{\prime},\lambda-2\delta_{c}}\right)\nonumber \\
 & +\frac{1}{2}\left(V_{\mu,\lambda+\delta_{c}}-V_{\mu,\lambda-\delta_{c}}\right)\delta_{\lambda^{\prime},\lambda}\nonumber \\
 & \left.\times\left(\delta_{\mu^{\prime},\mu+4\delta_{b}}-2(1+2\delta_{b}^{2}\gamma^{2})\delta_{\mu^{\prime},\mu}+\delta_{\mu^{\prime},\mu-4\delta_{b}}\right)\right],\label{eq:matrix-elmnt}
\end{align}
where $V_{\mu,\lambda}$ is the eigenvalue of the quantum volume operator
$\hat{V}$ in this basis. It is computed by using (\ref{eq:p-rep})
to represent the classical volume (\ref{eq:class-vol}), and then
acting it on this basis, 
\begin{equation}
\hat{V}\vert\mu,\lambda\rangle=V_{\mu,\lambda}\vert\mu,\lambda\rangle=2\pi\gamma^{3/2}\ell_{\text{Pl}}^{3}\vert\mu\vert\vert\lambda\vert^{1/2}\vert\mu,\lambda\rangle.
\end{equation}
Using this, the matrix element (\ref{eq:matrix-elmnt}) becomes

\begin{align}
\left\langle \mu_{n},\lambda_{n}\left|\hat{C}_{\textrm{S}}^{\left(\delta_{b},\delta_{c}\right)}\right|\mu_{n-1},\lambda_{n-1}\right\rangle = & \frac{2}{\gamma^{3}\ell_{\textrm{Pl}}^{2}}\left(\frac{1}{2\pi}\right)^{2}\int_{-\pi/p_{b}^{0}}^{\pi/p_{b}^{0}}db_{n}\int_{-\pi/p_{c}^{0}}^{\pi/p_{c}^{0}}dc_{n}\,\left\{ e^{-ib_{n}\left(p_{n}^{b}-p_{n-1}^{b}\right)-ic_{n}\left(p_{n}^{c}-p_{n-1}^{c}\right)}\phantom{\left(\frac{\sin^{2}(\delta b_{n})}{\delta_{b}^{2}}+\gamma^{2}\right)}\right. \nonumber\\ 
 & \left.\times\left[2V_{1}^{(n)}\frac{\sin(\delta_{b}b_{n})}{\delta_{b}}\frac{\sin(\delta_{c}c_{n})}{\delta_{c}}+V_{2}^{(n)}\left(\frac{\sin^{2}(\delta b_{n})}{\delta_{b}^{2}}+\gamma^{2}\right)\right]\right\} ,\label{eq:matrix-elmnt-2}
\end{align}
where
\begin{align}
V_{2}^{(n)}\coloneqq & \begin{cases}
\pi\gamma^{3/2}\ell_{\text{Pl}}^{3}\vert\mu_{n}\vert\frac{(\lambda_{n}+\delta_{c})^{1/2}-(\lambda_{n}-\delta_{c})^{1/2}}{\delta_{c}} & \lambda_{n}\geq\delta_{c}\\
\pi\gamma^{3/2}\ell_{\text{Pl}}^{3}\vert\mu_{n}\vert\frac{(\lambda_{n}+\delta_{c})^{1/2}-(\delta_{c}-\lambda_{n})^{1/2}}{\delta_{c}} & \vert\lambda_{n}\vert < \delta_{c}\\
\pi\gamma^{3/2}\ell_{\text{Pl}}^{3}\vert\mu_{n}\vert\frac{(-\lambda_{n}-\delta_{c})^{1/2}-(\delta_{c}-\lambda_{n})^{1/2}}{\delta_{c}} & \lambda_{n}\leq-\delta_{c}
\end{cases}\nonumber \\
= & 4\pi\gamma\ell_{\text{Pl}}^{2}\vert p_{n}^{b}\vert\frac{\left(\sqrt{\vert p_{n}^{c}+p_{c}^{0}\vert}-\sqrt{\vert p_{n}^{c}-p_{c}^{0}\vert}\right)}{p_{c}^{0}},\label{eq:aprox2-1}
\end{align}
and
\begin{align}
V_{1}^{(n)}\coloneqq & \begin{cases}
\pi\gamma^{3/2}\ell_{\text{Pl}}^{3}\vert\lambda_{n}\vert^{1/2} & \mu_{n}\geq\delta_{b}\\
\pi\gamma^{3/2}\ell_{\text{Pl}}^{3}\vert\lambda_{n}\vert^{1/2}\mu_{n}/\delta_{b} & \vert\mu_{n}\vert < \delta_{b}\\
-\pi\gamma^{3/2}\ell_{\text{Pl}}^{3}\vert\lambda_{n}\vert^{1/2} & \mu_{n}\leq-\delta_{b}
\end{cases}\nonumber \\
= & 4\pi\gamma\ell_{\text{Pl}}^{2}\vert p_{n}^{c}\vert^{\frac{1}{2}}\frac{\vert p_{n}^{b}+p_{b}^{0}\vert-\vert p_{n}^{b}+p_{b}^{0}\vert}{2p_{b}^{0}},
\end{align}
with
\begin{align}
p_{b}^{0}\coloneqq & \frac{1}{2}\gamma\ell_{\text{Pl}}^{2}\delta_{b}, & p_{c}^{0}\coloneqq & \gamma\ell_{\text{Pl}}^{2}\delta_{c}.\label{eq:p0s}
\end{align}
Substituting all these back into the short-time amplitude (\ref{eq:Exp-Pi-1})
yields
\begin{align}
\left\langle \mu_{n},\lambda_{n}\left|e^{-i\epsilon\alpha \hat{C}_{\textrm{S}}^{\left(\delta_{b},\delta_{c}\right)}}\right|\mu_{n-1},\lambda_{n-1}\right\rangle = & \int_{-\pi/p_{b}^{0}}^{\pi/p_{b}^{0}}db_{n}\int_{-\pi/p_{c}^{0}}^{\pi/p_{c}^{0}}dc_{n}\,\left\{ e^{-ib_{n}\left(p_{n}^{b}-p_{n-1}^{b}\right)-ic_{n}\left(p_{n}^{c}-p_{n-1}^{c}\right)}\right.\nonumber \\
 & \left.\times\left(1-i\epsilon\alpha \tilde{C}\left(p_{n}^{b},p_{n}^{c},b_{n},c_{n}\right)\right)\right\} +\mathcal{O}(\epsilon^{2})\nonumber \\
= & \int_{-\pi/p_{b}^{0}}^{\pi/p_{b}^{0}}db_{n}\int_{-\pi/p_{c}^{0}}^{\pi/p_{c}^{0}}dc_{n}\,\left\{ e^{-ib_{n}\left(p_{n}^{b}-p_{n-1}^{b}\right)-ic_{n}\left(p_{n}^{c}-p_{n-1}^{c}\right)-i\epsilon\alpha \tilde{C}\left(p_{n}^{b},p_{n}^{c},b_{n},c_{n}\right)}\right\} \nonumber \\
 & +\mathcal{O}(\epsilon^{2}),
\end{align}
where \textcolor{red}{}
\begin{equation}
\tilde{C}(p_{n}^{b},p_{n}^{c},b_{n},c_{n})=\frac{2}{\gamma^{3}\ell_{\textrm{Pl}}^{2}}\left[2V_{1}^{(n)}\frac{\sin(\delta_{b}b_{n})}{\delta_{b}}\frac{\sin(\delta_{c}c_{n})}{\delta_{c}}+V_{2}^{(n)}\left(\frac{\sin^{2}(\delta b_{n})}{\delta_{b}^{2}}+\gamma^{2}\right)\right].
\end{equation}
The amplitude (\ref{eq:Kernel-1}), is the multiplication of these `short-time" ampltudes, 
\begin{align*}
A\left(\mu_{f},\lambda_{f},\mu,\lambda_{i} \right)= & \int d\alpha\prod_{n=1}^{\mathcal{N}}\sum_{\mu_{n},\lambda_{n}\in\Gamma}\int_{-\pi/p_{b}^{0}}^{\pi/p_{b}^{0}}db_{n}\int_{-\pi/p_{c}^{0}}^{\pi/p_{c}^{0}}dc_{n}\bigg\{\\
 & \left.\exp\left(-i\left[\epsilon\sum_{n=1}^{\mathcal{N}}\frac{b_{n}\left(p_{n}^{b}-p_{n-1}^{b}\right)+c_{n}\left(p_{n}^{c}-p_{n-1}^{c}\right)}{\epsilon}+\epsilon\alpha \sum_{n=1}^{\mathcal{N}}\tilde{C}\left(p_{n}^{b},p_{n}^{c},b_{n},c_{n}\right)\right]\right)\right\} \\
 & +\mathcal{O}(\epsilon^{2}).
\end{align*}
The first term in the exponential can be written as
\begin{equation}
\sum_{n=1}^{\mathcal{N}}b_{n}\left(p_{n}^{b}-p_{n-1}^{b}\right)+c_{n}\left(p_{n}^{c}-p_{n-1}^{c}\right)=\textrm{B.T.}-\sum_{n=1}^{\mathcal{N}-1}\left(b_{n+1}-b_{n}\right)p_{n}^{b}+\left(c_{n+1}-c_{n}\right)p_{n}^{c},
\end{equation}
where B.T. is the boundary term
\begin{equation}
\textrm{B.T.}=b_{\mathcal{N}}p_{\mathcal{N}}^{b}-b_{1}p_{0}^{b}+c_{\mathcal{N}}p_{\mathcal{N}}^{c}-c_{1}p_{0}^{c}.
\end{equation}
Then, the amplitude becomes
\begin{align*}
A\left(\mu_{f},\lambda_{f},\mu,\lambda_{i} \right)= &\int d\alpha  \prod_{n=1}^{\mathcal{N}}\sum_{\mu_{n},\lambda_{n}\in\Gamma}\int_{-\pi/p_{b}^{0}}^{\pi/p_{b}^{0}}db_{n}\int_{-\pi/p_{c}^{0}}^{\pi/p_{c}^{0}}dc_{n}\bigg\{\\
 & \left.\exp\left(i\left[\epsilon\sum_{n=1}^{\mathcal{N}}\frac{b_{n+1}-b_{n}}{\epsilon}p_{n}^{b}+\frac{c_{n+1}-c_{n}}{\epsilon}p_{n}^{c}-\epsilon\alpha \sum_{n=1}^{\mathcal{N}}\tilde{C}\left(p_{n}^{b},p_{n}^{c},b_{n},c_{n}\right)+\textrm{B.T.}\right]\right)\right\} \\
 & +\mathcal{O}(\epsilon^{2}).
\end{align*}
Finally taking the limit $\mathcal{N}\rightarrow\infty$ such that
$\mathcal{N}\epsilon=1$, we arrive at
\begin{align}
A(\mu_{f},\lambda_{f},\mu,\lambda_{i} )=\int d\alpha \int Db\int Dc\exp & \left\{ i\int_{t_{i}}^{t_{f}}\text{d}t\left[p_{b}\dot{b}+p_{c}\dot{c}-\alpha C_{\textrm{S-eff}}^{\left(\delta_{b},\delta_{c}\right)}\left(p_{b}(t),p_{c}(t),b(t),c(t)\right)\right]\right.\nonumber \\
 & +\underbrace{b_{f}p_{f}^{b}-b_{i}p_{i}^{b}+c_{f}p_{f}^{c}-c_{i}p_{i}^{c}}_{\textrm{B.T.}}\bigg\}
\end{align}
where in the limit taken, $\epsilon\sum_{n=1}^{\mathcal{N}-1}\rightarrow\int\text{d}t$,
and $p_{n}^{c},p_{n}^{b}\rightarrow p_{c}(t),p_{b}(t)$, respectively,
and 
\begin{equation}
\int Db\int Dc=\lim_{\mathcal{N}\rightarrow\infty}\prod_{n=1}^{\mathcal{N}}\sum_{\mu_{n},\lambda_{n}\in\Gamma}\int_{-\pi/p_{b}^{0}}^{\pi/p_{b}^{0}}db_{n}\int_{-\pi/p_{c}^{0}}^{\pi/p_{c}^{0}}dc_{n}.
\end{equation}
To transform the integrals from a bounded interval to the whole real line we use the Jacobi identity for periodic functions (Eq. (3.13) in \cite{Ashtekar:2010gz}). From this, one can read off the effective Hamiltonian constraint from
the path integral representation of the kernel as\textcolor{red}{}
\begin{equation}
C_{\textrm{S-eff}}^{\left(\delta_{b},\delta_{c}\right)}\left(p_{b},p_{c},p_{b},p_{c}\right)=-\frac{2}{\gamma^{3}\ell_{\textrm{Pl}}^{2}}\left[2V_{1}\left(p_{b},p_{c}\right)\frac{\sin\left(\delta_{b}b\right)}{\delta_{b}}\frac{\sin\left(\delta_{c}c\right)}{\delta_{c}}+V_{2}\left(p_{b},p_{c}\right)\left(\frac{\sin^{2}\left(\delta_{b}b\right)}{\delta_{b}^{2}}+\gamma^{2}\right)\right].\label{eq:efect-H-PI}
\end{equation}
Here
\begin{equation}
V_{1}\left(p_{b},p_{c}\right)=\lim_{p_{n}^{b},p_{n}^{c}\rightarrow p_{b}(t),p_{c}(t)}V_{1}^{(n)}=4\pi\gamma\ell_{\text{Pl}}^{2}\beta_{1}\left(p_{b},p_{b}^{0}\right) p_{c}^{\frac{1}{2}},\label{eq:Vb-beta1}
\end{equation}
with
\begin{equation}
\beta_{1}\left(p_{b},p_{b}^{0}\right)=\frac{\vert p_{b}+p_{b}^{0}\vert-\vert p_{b}-p_{b}^{0}\vert}{2p_{b}^{0}}.\label{eq:beta1}
\end{equation}
In the same manner 
\begin{equation}
V_{2}\left(p_{b},p_{c}\right)=\lim_{p_{n}^{b},p_{n}^{c}\rightarrow p_{b}(t),p_{c}(t)}V_{2}^{(n)}=4\pi\gamma\ell_{\text{Pl}}^{2}\beta_{2}\left(p_{c},p_{c}^{0}\right)\frac{ p_{b}}{p_{c}^{\frac{1}{2}}},\label{eq:Vc-beta2}
\end{equation}
in which
\begin{equation}
\beta_{2}\left(p_{c},p_{c}^{0}\right)=p_{c}^{\frac{1}{2}}\frac{\left(\sqrt{\vert p_{c}+p_{c}^{0}\vert}-\sqrt{\vert p_{c}-p_{c}^{0}\vert}\right)}{p_{c}^{0}}.\label{eq:beta2}
\end{equation}
Note that due to the form of $p_c,\, p_{c}^{0}, \, p_b,\, p_{b}^{0}$, under a rescaling $L_{0}\rightarrow\xi L_{0}$ we get
\begin{align}
b\to & b^{\prime}=b & c\to & c^{\prime}=\xi c\\
p_{b}\to & p_{b}^{\prime}=\xi p_{b} & p_{c}\to & p_{c}^{\prime}=p_{c}\\
\delta_{b}= & \frac{\sqrt{\Delta}}{r_{0}}\to\delta_{b}^{\prime}=\delta_{b} & \delta_{c}= & \frac{\sqrt{\Delta}}{L_{0}}\to\delta_{c}^{\prime}=\frac{\delta_{c}}{\xi}\\
\delta_{b}l_{b}= & \delta_{b}r_{0} & \delta_{c}l_{c}= & \delta_{c}L_{0}\to\delta_{c}\xi L_{0}\\
p_{b}^{0}= & \frac{1}{2}\gamma\ell_{\text{Pl}}^{2}\delta_{b}\to p_{b}^{0\prime}=p_{b}^{0} & p_{c}^{0}= & \gamma\ell_{\text{Pl}}^{2}\delta_{c}\to p_{c}^{0\prime}=\frac{p_{c}^{0}}{\xi}
\end{align}
Using these and the form of $\beta_{1}\left(p_{b},p_{b}^{0}\right)$ and $\beta_{2}\left(p_{c},p_{c}^{0}\right)$, we obtain the rescaling for these expressions as
\begin{align}
\beta_{1}\left(p_{b}^{\prime},p_{b}^{0\prime}\right)= & \frac{\vert p_{b}^{\prime}+p_{b}^{0\prime}\vert-\vert p_{b}^{\prime}-p_{b}^{0\prime}\vert}{2p_{b}^{0\prime}}\nonumber \\
= & \frac{\vert\xi p_{b}+p_{b}^{0}\vert-\vert\xi p_{b}-p_{b}^{0}\vert}{2p_{b}^{0}},\label{rescal-beta1}
\end{align}
and
\begin{align}
\beta_{2}\left(p_{c}^{\prime},p_{c}^{0\prime}\right)= & p_{c}^{\prime\frac{1}{2}}\frac{\left(\sqrt{\vert p_{c}^{\prime}+p_{c}^{0\prime}\vert}-\sqrt{\vert p_{c}^{\prime}-p_{c}^{0\prime}\vert}\right)}{p_{c}^{0\prime}}\nonumber \\
= & p_{c}^{\frac{1}{2}}\frac{\left(\sqrt{\vert p_{c}+\frac{p_{c}^{0}}{\xi}\vert}-\sqrt{\vert p_{c}-\frac{p_{c}^{0}}{\xi}\vert}\right)}{\frac{p_{c}^{0}}{\xi}},\label{rescal-beta2}
\end{align}
and hence neither of the functions $\beta_{1}$ and $\beta_{2}$ are
invariant.

\begin{figure}
\noindent \centering{}\includegraphics[scale=.35]{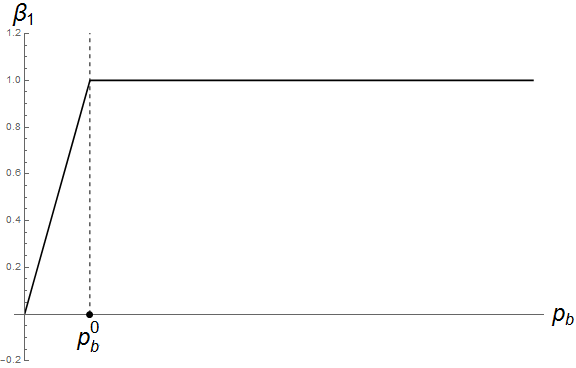}~~~~~~~~~~~\includegraphics[scale=.35]{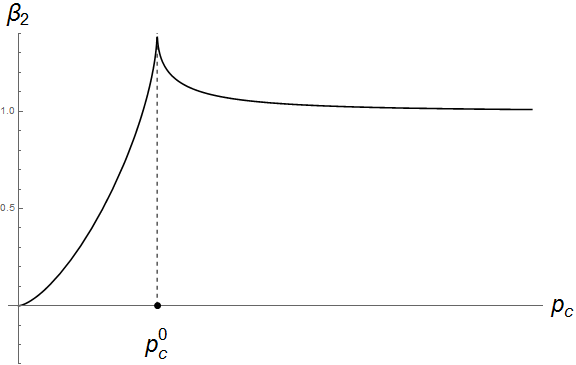}\caption{The functions $\beta_{1}$ and $\beta_{2}$, with $p_{b}=1=p_{c}$.
It is seen that for $p_{b}>p_{b}^{0}$ and $p_{c}\gg p_{c}^{0}$, they behave as $\beta_{1},\beta_{2}\to1$.\label{fig:betas}}
\end{figure}

With introduction of $\beta_{1}$ and $\beta_{2}$, the effective
Hamiltonian (\ref{eq:efect-H-PI}) times the lapse function, $\frac{N}{16\pi G}$,
is written as\textcolor{red}{}
\begin{align}
C_{\textrm{S-eff}}= & \frac{N}{16\pi G}C_{\textrm{S-eff}}^{\left(\delta_{b},\delta_{c}\right)}\nonumber \\
= & -\frac{N}{2G\gamma^{2}}\left[2\beta_{1}\left(p_{b},p_{b}^{0}\right) p_{c}^{\frac{1}{2}}\frac{\sin\left(\delta_{b}b\right)}{\delta_{b}}\frac{\sin\left(\delta_{c}c\right)}{\delta_{c}}+\beta_{2}\left(p_{c},p_{c}^{0}\right)\frac{ p_{b}}{p_{c}^{\frac{1}{2}}}\left(\frac{\sin^{2}\left(\delta_{b}b\right)}{\delta_{b}^{2}}+\gamma^{2}\right)\right].\label{eq:H-eff-beta}
\end{align}
This effective Hamiltonian resembles the ones that have been suggested
in previous works, with the important difference of incorporating
further inverse triad quantum corrections, encoded in functions $\beta_{1}$
and $\beta_{2}$, as can be seen from (\ref{eq:beta1}) and (\ref{eq:beta2}). The profile of these functions are plotted in Fig. \ref{fig:betas}. 

For a generic lapse function $N,$ this effective Hamiltonian leads
to the equations of motion, $\dot{F}=\left\{ F,C_{\textrm{S-eff}}\right\} $,
that read\textcolor{red}{}
\begin{align}
\dot{b}= & -\frac{1}{2\gamma}\left[2\beta_{1}\left(p_{b},p_{b}^{0}\right) p_{c}^{\frac{1}{2}}\frac{\sin\left(\delta_{b}b\right)}{\delta_{b}}\frac{\sin\left(\delta_{c}c\right)}{\delta_{c}}\left\{ b,N\right\} \right.\nonumber \\
 & +\beta_{2}\left(p_{c},p_{c}^{0}\right)\frac{1}{p_{c}^{\frac{1}{2}}}\left(\frac{\sin^{2}\left(\delta_{b}b\right)}{\delta_{b}^{2}}+\gamma^{2}\right)\left( p_{b}\left\{ b,N\right\} +N\left(p_{b}\right)\right)\nonumber \\
 & \left.+2N\frac{\partial\beta_{1}\left(p_{b},p_{b}^{0}\right)}{\partial p_{b}} p_{c}^{\frac{1}{2}}\frac{\sin\left(\delta_{b}b\right)}{\delta_{b}}\frac{\sin\left(\delta_{c}c\right)}{\delta_{c}}\right],\label{eq:gen-eqm-b}\\
\dot{c}= & -\frac{1}{\gamma}\left[\beta_{1}\left(p_{b},p_{b}^{0}\right) p_{c}^{\frac{1}{2}}\frac{\sin\left(\delta_{b}b\right)}{\delta_{b}}\frac{\sin\left(\delta_{c}c\right)}{\delta_{c}}\left(2\left\{ c,N\right\} +\frac{N}{p_{c}}\right)\right.\nonumber \\
 & \left.+\beta_{2}\left(p_{c},p_{c}^{0}\right)\frac{p_{b}}{p_{c}^{\frac{1}{2}}}\left(\left\{ c,N\right\} -\frac{N}{2p_{c}}\right)\left(\frac{\sin^{2}\left(\delta_{b}b\right)}{\delta_{b}^{2}}+\gamma^{2}\right)+N\frac{ p_{b}}{p_{c}^{\frac{1}{2}}}\frac{\partial\beta_{2}\left(p_{c},p_{c}^{0}\right)}{\partial p_{c}}\left(\frac{\sin^{2}\left(\delta_{b}b\right)}{\delta_{b}^{2}}+\gamma^{2}\right)\right],\\
\dot{p}_{b}= & -\frac{1}{2\gamma}\left[2\beta_{1}\left(p_{b},p_{b}^{0}\right) p_{c}^{\frac{1}{2}}\frac{\sin\left(\delta_{c}c\right)}{\delta_{c}}\left(\frac{\sin\left(\delta_{b}b\right)}{\delta_{b}}\left\{ p_{b},N\right\} -N\cos\left(\delta_{b}b\right)\right)\right.\nonumber \\
 & \left.+\beta_{2}\left(p_{c},p_{c}^{0}\right)\frac{ p_{b}}{p_{c}^{\frac{1}{2}}}\left(-2N\frac{\sin\left(\delta_{b}b\right)}{\delta_{b}}\cos\left(\delta_{b}b\right)+\left(\frac{\sin^{2}\left(\delta_{b}b\right)}{\delta_{b}^{2}}+\gamma^{2}\right)\left\{ p_{b},N\right\} \right)\right],\\
\dot{p}_{c}= & -\frac{1}{\gamma}\left[2\beta_{1}\left(p_{b},p_{b}^{0}\right) p_{c}^{\frac{1}{2}}\frac{\sin\left(\delta_{b}b\right)}{\delta_{b}}\left(\frac{\sin\left(\delta_{c}c\right)}{\delta_{c}}\left\{ p_{c},N\right\} -N\cos\left(\delta_{c}c\right)\right)\right.\nonumber \\
 & \left.+\beta_{2}\left(p_{c},p_{c}^{0}\right)\frac{ p_{b}}{p_{c}^{\frac{1}{2}}}\left(\frac{\sin^{2}\left(\delta_{b}b\right)}{\delta_{b}^{2}}+\gamma^{2}\right)\left\{ p_{c},N\right\} \right].\label{eq:gen-eqm-pb}
\end{align}
Note that although the derivative of $\beta_{i}$ are not well-defined
at the kink at $p^{0}$, nevertheless given the existence of a minimal
radius at bounce, $p=p^{0}$ will not happen and one can use the derivatives of $\beta_{i}$ in the above equations safely.

These equations will help us clarify some of the differences of our
results from the previous ones, in the next sections.

\section{Issues raised by the new corrections\label{sec:Issues-new-correc}}

As mentioned earlier, in this model, a number of differences arise
due to the presence of further inverse triad corrections that we have managed to compute through the path integral method. To further
highlight these differences, and also to be able to compare our results
with some of the previous works, we need to specify a specific lapse
$N$, which is needed to write the explicit equations of motion in
a certain frame. One such choice takes us to the 
Hamiltonian in \cite{Corichi:2015xia} in the limit of not considering these additional inverse
triad quantum corrections, {\em i.e.} when $\beta_{1},\beta_{2}\to1$, is
\begin{equation}
N^{(1)}=\frac{\gamma\delta_{b}p_{c}^{\frac{1}{2}}}{\sin\left(\delta_{b}b\right)},\label{eq:N(1)}
\end{equation}
for which the effective Hamiltonian becomes
\begin{equation}
C_{\textrm{eff}}^{1}=-\frac{1}{2G\gamma}\left[2\beta_{1}\left(p_{b},p_{b}^{0}\right) p_{c}\frac{\sin\left(\delta_{c}c\right)}{\delta_{c}}+\beta_{2}\left(p_{c},p_{c}^{0}\right) p_{b}\left(\frac{\sin\left(\delta_{b}b\right)}{\delta_{b}}+\gamma^{2}\frac{\delta_{b}}{\sin\left(\delta_{b}b\right)}\right)\right].\label{eq:HE1}
\end{equation}
This lapse, which at the classical level is $N=\frac{\gamma\sqrt{p_{c}}}{b}$, is quite useful because it decouples the classical equations of motion for $b$ and $p_b$ from those of $c$ and $p_c$, and hence simplifies many of the associated analyses. Thus, in order to be able to compare our effective results with their classical counter parts, we need to use the same lapse, now in its effective form where $b\to \sin(\delta_b b)/b$.   

It is clearly seen that this Hamiltonian matches that in \cite{Corichi:2015xia}
except for the presence of additional inverse triad corrections $\beta_{1},\beta_{2}$,
while they will match exactly for $\beta_{1},\beta_{2}\to1$.

The equations of motion corresponding to this effective Hamiltonian
can be derived by using the lapse (\ref{eq:N(1)}) in the equations
of motion (\ref{eq:gen-eqm-b})-(\ref{eq:gen-eqm-pb}) which yields

\begin{align}
\dot{b}= & -\frac{1}{2}\left[2 p_{c}\frac{\sin\left(\delta_{c}c\right)}{\delta_{c}}\frac{\partial\beta_{1}\left(p_{b},p_{b}^{0}\right)}{\partial p_{b}}+\beta_{2}\left(p_{c},p_{c}^{0}\right)p_{b}\left(\frac{\sin\left(\delta_{b}b\right)}{\delta_{b}}+\gamma^{2}\frac{\delta_{b}}{\sin\left(\delta_{b}b\right)}\right)\right],\\
\dot{c}= & -\left[2\beta_{1}\left(p_{b},p_{b}^{0}\right)\frac{\sin\left(\delta_{c}c\right)}{\delta_{c}}+p_{b}\frac{\partial\beta_{2}\left(p_{c},p_{c}^{0}\right)}{\partial p_{c}}\left(\frac{\sin\left(\delta_{b}b\right)}{\delta_{b}}+\gamma^{2}\frac{\delta_{b}}{\sin\left(\delta_{b}b\right)}\right)\right]\\
\dot{p}_{b}= & \frac{1}{2}\beta_{2}\left(p_{c},p_{c}^{0}\right)p_{b}\cos\left(\delta_{b}b\right)\left(1-\gamma^{2}\frac{\delta_{b}^{2}}{\sin^{2}\left(\delta_{b}b\right)}\right),\\
\dot{p}_{c}= & 2\beta_{1}\left(p_{b},p_{b}^{0}\right)p_{c}\cos\left(\delta_{c}c\right).\label{eq:minimop}
\end{align}
The lapse (\ref{eq:N(1)}), however, is not the only choice for which
the effective Hamiltonian will be the same as the Hamiltonian \cite{Corichi:2015xia},
in the limit $\beta_{1},\beta_{2}\to1$. In fact any other lapse functions
$N^{(2)}$, such that $\lim_{\beta_{1},\beta_{2}\rightarrow1}N^{(1)}=\lim_{\beta_{1},\beta_{2}\rightarrow1}N^{(2)}$,
will do, for which $N^{(2)}=\frac{\gamma\delta_{b}\sqrt{p_{c}}}{\beta_{2}\sin\left(\delta_{b}b\right)}$ is an example.

One of the differences due to the presence of the $\beta_{1},\beta_{2}$
functions is related to the minimum value of $p_{c}$. In classical
theory, at the horizon, $p_{c}=4G^{2}M^{2}$, and by approaching the
singularity, $p_{b}\to0$ and $p_{c}\to0$, as stated in  \ref{eq:t-singular}. But,
in the effective theory, due to the discreteness of the geometry,
both of these functions bounce at the singularity and their minimum
is related to the minimum length. In our
model however, regardless of the lapse function, this minimum is different
from what has been computed in previous works. To get this minimum
value, we first note that
\begin{equation}
Q=\frac{\sin(\delta_{c}c)}{\delta_{c}}\frac{p_{c}}{\beta_{2}\left(p_{c},p_{c}^{0}\right)},\label{eq:Q-def}
\end{equation}
is a weak Dirac observable  since
\begin{equation}
\dot{Q}=\left\{ Q,C_{\textrm{S-eff}}^{\left(\delta_{b},\delta_{c}\right)}\right\} \approx \frac{2G\gamma\cos\left(\delta_{c}c\right)p_{c}}{\beta_{2}^{2}\left(p_{c},p_{c}^{0}\right)}\frac{\partial\beta_{2}\left(p_{c},p_{c}^{0}\right)}{\partial p_{c}}C_{\textrm{S-eff}}^{\left(\delta_{b},\delta_{c}\right)}\approx0.
\end{equation}
On the other hand we can see from (\ref{eq:minimop}) that the extremum
of $p_{c}$ happens for $\cos\left(\delta_{c}c\right)=0$ or $\sin\left(\delta_{c}c\right)=\pm1$.
Notice that we are not claiming  $\sin\left(\delta_{c}c\right)=\pm1$ leads to the minimum of $Q$. Rather, we are deducing a minimum value for $p_c$ solely from \eqref{eq:minimop} and independently of \eqref{eq:Q-def} without any reference to $Q$. Using the value $\sin\left(\delta_{c}c\right)=+1$ in (\ref{eq:Q-def}) yields 
\begin{equation}
p_{c(\textrm{min})}=Q\delta_{c}\beta_{2}\left(p_{c(\textrm{min})},p_{c}^{0}\right).\label{eq:pc-min-Q}
\end{equation}
After solving for $p_{c\text{(min)}}$
and using (\ref{eq:p0s}),  one finds the minimum value to be
\begin{equation}
p_{c\text{(min)}}=\frac{Q\Delta^{\frac{1}{2}}}{L_{0}}\frac{1}{\sqrt{1-\left(\frac{\gamma\ell_{\text{Pl}}^{2}}{2Q}\right)^{2}}},\,\,\,\,\,\,\,p_{c}>p_{c}^{0}.\label{eq:pc-min}
\end{equation}
The weak Dirac observable similar to $Q$ in \cite{Corichi:2015xia} is $Q^{\textrm{C-S}}=\frac{\sin(\delta_{c}c)}{\delta_{c}}p_{c}$,
which makes their minimum value $p_{c(\textrm{min})}^{\textrm{C-S}}=Q^{\textrm{C-S}}\delta_{c}$.
%From this and (\ref{eq:pc-min-Q}), we get
%\begin{equation}
%\frac{p_{c(\textrm{min})}}{\beta_{2}\left(p_{c(\textrm{min})},p_{c}^{0}\right)}=p_{c(\textrm{min})}^{\textrm{C-S}}\frac{Q}{Q^{\textrm{C-S}}}
%\end{equation}
%which after solving for $p_{c(\textrm{min})}$ yields
%\begin{equation}
%p_{c(\textrm{min})}=\pm\left(\frac{Q}{Q^{\textrm{C-S}}}\right)^{2}\frac{p_{c(\textrm{min})}^{\textrm{C-S}}}{\sqrt{\left(\frac{Q}{Q^{\textrm{C-S}}}\right)^{2}-\left(\frac{p_{c}^{0}}{2p_{c(\textrm{min})}^{\textrm{C-S}}}\right)^{2}}},\,\,\,\,\,\,p_{c}>p_{c}^{0}.
%\end{equation}
The value of the constant of motion $Q$ can be determined by the initial conditions of the equations of motion. In \cite{Corichi:2015xia} the initial conditions for the equations of motion are identified such that when solving, they lead directly to the Schwarzschild metric. We use the same initial values as \cite{Corichi:2015xia}, so that in the classical limit we directly obtain the Schwarzchild metric. Hence, if in the above equation, we take  $Q$ equal to the constant of motion in \cite{Corichi:2015xia}
$Q^{\textrm{C-S}}=\gamma L_{0}GM$, we will get
\begin{equation}
p_{c(\textrm{min})}=\gamma GM\sqrt{\Delta}\frac{1}{\sqrt{1-\left(\frac{\ell_{\text{Pl}}^{2}}{2GML_{0}}\right)^{2}}},\,\,\,\,\,\,p_{c}>p_{c}^{0}.
\end{equation}
This is the value of the $p_{c}$ at time of the bounce, and has
a pure quantum origin such that for $G\hbar\to0$, one gets $\Delta\to0$
and thus $p_{c(\textrm{min})}\to0$. Clearly this new value for minimum
of $p_{c}$ is different from what computed in \cite{Corichi:2015xia}
by a factor of $\frac{1}{\sqrt{1-\frac{1}{4}\left(\frac{\ell_{\text{Pl}}^{2}}{GML_{0}}\right)^{2}}}$,
but, it depends on the auxiliary parameter $L_{0}$. This dependence
on $L_{0}$ or its rescaling $L_{0}\rightarrow\xi L_{0}$, shows itself
in several places. Let us consider for instance a congruence of geodesic observers and their corresponding shear and expansion.Their dependence upon $L_0$  is an undesirable effect, since the physical results should
not depend on the an auxiliary variable or its rescaling. This 
can be traced back to the presence of the new corrections $\beta_{1}$
and $\beta_{2}$. To see this in a more concrete way, we first consider
the expansion $\theta$. For a generic lapse it can be written as
\begin{equation}
\theta=\frac{\dot{p}_{b}}{Np_{b}}+\frac{\dot{p}_{c}}{2Np_{c}},
\end{equation}
which for $N=1$ (and on constraint surface) turns out to be
\begin{align}
\theta= & \frac{1}{\gamma}\left\{ \beta_{1}\left(p_{b},p_{b}^{0}\right) p_{c}^{\frac{1}{2}}\left[\frac{\sin\left(\delta_{c}c\right)\cos\left(\delta_{b}b\right)}{p_{b}\delta_{c}}+\frac{\sin\left(\delta_{b}b\right)\cos\left(\delta_{c}c\right)}{p_{c}\delta_{b}}\right]\right.\nonumber \\
 & \left.+\beta_{2}\left(p_{c},p_{c}^{0}\right)\frac{1}{ p_{c}^{\frac{1}{2}}}\frac{\sin\left(\delta_{b}b\right)\cos\left(\delta_{b}b\right)}{\delta_{b}}\right\} .\label{eq:theta-orig-N1}
\end{align}
%From appendix \ref{Appx:Transf}
From the transformation properties of the objects involved, it can be seen that although all the
combination $\delta_{b}b,\,\delta_{c}c,\,p_{b}\delta_{c},\,p_{c}\delta_{b}$
in the terms above are invariant under a rescaling $L_{0}\rightarrow\xi L_{0}$,
the expansion $\theta$ itself is not, precisely because the presence
and noninvariance of $\beta_{1}\left(p_{b},p_{b}^{0}\right)$ and
$\beta_{2}\left(p_{c},p_{c}^{0}\right)$.

We can also compute shear $\sigma^{2}$ 
\begin{equation}
\sigma^{2}=\frac{1}{3}\left(-\frac{\dot{p}_{b}}{Np_{b}}+\frac{\dot{p}_{c}}{Np_{c}}\right)^{2}
\end{equation}
to see that for $N=1$ we get
\begin{align}
\sigma^{2}= & \frac{1}{3\gamma^{2}}\left\{ \beta_{1}\left(p_{b},p_{b}^{0}\right) p_{c}^{\frac{1}{2}}\left[2\frac{\sin\left(\delta_{b}b\right)\cos\left(\delta_{c}c\right)}{p_{c}\delta_{b}}-\frac{\sin\left(\delta_{c}c\right)\cos\left(\delta_{b}b\right)}{p_{b}\delta_{c}}\right]\right.\nonumber \\
 & \left.-\beta_{2}\left(p_{c},p_{c}^{0}\right)\frac{\left(p_{b}\right)}{p_{c}^{\frac{1}{2}}}\frac{\sin\left(\delta_{b}b\right)\cos\left(\delta_{b}b\right)}{\delta_{b}}\right\} ^{2}.\label{eq:sigma-orig-N1}
\end{align}
Again in the terms above, the only noninvariant parts under rescalings
are $\beta_{1}\left(p_{b},p_{b}^{0}\right)$ and $\beta_{2}\left(p_{c},p_{c}^{0}\right)$ as was discussed in Eqs. \eqref{rescal-beta1} and \eqref{rescal-beta2}.
Thus a strong hint is that a solution that renders both $\beta_{1}$
and $\beta_{2}$, invariant under rescalings, will resolve all of
the above issues. Finding such a solution is the subject of the next
section.
\begin{figure}
\noindent \centering{}\includegraphics[scale=.5]{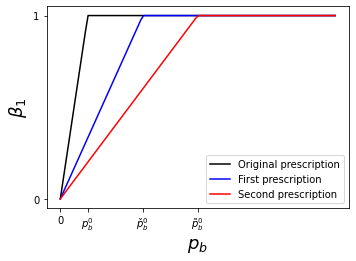}~~~~~~~~~~~\includegraphics[scale=.5]{beta1NEW.eps}\caption{Comparison of the the original $\beta_{1},\beta_{2}$ with the new
functions $\tilde{\beta}_{1},\tilde{\beta}_{2}$ of the first prescription
and $\check{\beta}_{1},\check{\beta}_{2}$ of the second prescription.\label{fig:betas12}}
\end{figure}

\section{Proposals to deal with the issues\label{sec:Prescriptions}}

The above observations together with the detailed forms of $\beta_{1}$
and $\beta_{2}$, suggest that some sort of interchanging $p_{b}^{0}\leftrightarrow p_{c}^{0}$
can fix the problem, either just in $\beta_{2}$, or in both $\beta_{1}$
and $\beta_{2}$. At a first glance, it seems that this can be achieved
by interchanging $\delta_{b}\leftrightarrow\delta_{c}$, but this
has two problems: it is not clear if there are restrictions in doing
so, and more importantly, although it fixes the invariance problem
in $\beta_{2}$, it makes almost every other terms in $\theta$ and
$\sigma^{2}$ noninvariant. So we should find a way of interchanging
$\delta_{b}\leftrightarrow\delta_{c}$ that only results in an interchange
$p_{b}^{0}\leftrightarrow p_{c}^{0}$, but does not lead to any modifications
or interchange of $\delta_{b}\leftrightarrow\delta_{c}$ outside $p_{b}^{0}$
and $p_{c}^{0}$. In other words, it should only affect triad corrections
but not the holonomy ones. By looking at (\ref{eq:H-constr}), (\ref{eq:class-curvatr}),
and (\ref{eq:thiemann-trick}), we notice that this can be achieved
by some sort of interchange of $\delta_{b}$ and $\delta_{c}$ but
just in the Thiemann's formula (\ref{eq:thiemann-trick}) (which is
allowed classically), and not in the computations of the curvature
in (\ref{eq:class-curvatr}).Given this observation, we make two proposals which work particularly well at or near the singularity,
and are explained in the following sections. 

\subsection{Proposal one: $\delta_{b}\leftrightarrow\delta_{c}$ in Thiemann's formula}

\begin{figure}
\noindent \begin{centering}
\includegraphics[scale=.6]{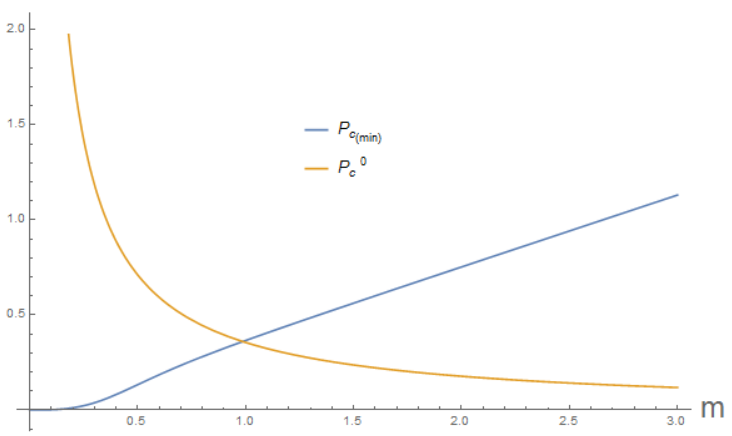}
\par\end{centering}
\caption{The effects of the corrections are relevant when $p_{c(\textrm{min})}$ is comparable
to, or smaller than $p_{c}^{0}$. In our proposal, this happens for $M_{B}\approx M_{\textrm{Pl}}$.\label{masacrit}}
\end{figure}

The first proposal is to mutually interchange, $r_0$ and $L_0$ in $\delta_{b}$,$\delta_{c}$ amounting to the exchange  $\delta_{b}\leftrightarrow\delta_{c}$
only in Thiemann's formula (\ref{eq:thiemann-trick}) which can
always be done classically. This
only affects the inverse triad correction by essentially interchanging
$p_{b}^{0}\leftrightarrow\frac{1}{2}p_{c}^{0}$, while not altering
the terms that depend on holonomy corrections. Consequently $\delta_{b}$
and $\delta_{c}$ remain the same whenever they appear outside $p_{b}^{0}$
or $p_{c}^{0}$. As we will see, this has several desired consequences.
After such an interchange $\delta_{b}\leftrightarrow\delta_{c}$,
the quantities $p_{b}^{0},\,p_{c}^{0},\,\beta_{1}$ and $\beta_{2}$
are replaced by their new versions denoted by a tilde as
\begin{align}
\tilde{p}_{b}^{0}= & \frac{1}{2}\gamma\ell_{\text{Pl}}^{2}\delta_{c}=\frac{1}{2}p_{c}^{0},\label{eq:pb0-tilde}\\
\tilde{p}_{c}^{0}= & \gamma\ell_{\text{Pl}}^{2}\delta_{b}=2p_{b}^{0},\label{eq:pc0-tilde}\\
\tilde{\beta}_{1}\left(p_{b},\tilde{p}_{b}^{0}\right)= & \frac{\vert p_{b}+\tilde{p}_{b}^{0}\vert-\vert p_{b}-\tilde{p}_{b}^{0}\vert}{2\tilde{p}_{b}^{0}},\label{eq:beta-1-tilde-1st-Pr}\\
\tilde{\beta}_{2}\left(p_{c},\tilde{p}_{c}^{0}\right)= & p_{c}^{\frac{1}{2}}\frac{\left(\sqrt{\vert p_{c}+\tilde{p}_{c}^{0}\vert}-\sqrt{\vert p_{c}-\tilde{p}_{c}^{0}\vert}\right)}{\tilde{p}_{c}^{0}}.\label{eq:beta-2-tilde-1st-Pr}
\end{align}
The profile of these new function $\tilde{\beta}_{1},\tilde{\beta}_{2}$
in comparison with the original $\beta_{1},\beta_{2}$ and the new
function from the second prescription $\check{\beta}_{1},\check{\beta}_{2}$
(see next section) can be seen in Fig. \ref{fig:betas12}. Considering (\ref{eq:delta-Delta}), we see that under a rescaling
$L_{0}\rightarrow\xi L_{0}$, one gets
\begin{align}
\tilde{p}_{b}^{0}\to\tilde{p}_{b}^{0\prime}= & \frac{\tilde{p}_{b}^{0}}{\xi},\\
\tilde{p}_{c}^{0}\to\tilde{p}_{c}^{0\prime}= & \tilde{p}_{c}^{0},\\
\tilde{\beta}_{1}\left(p_{b},\tilde{p}_{b}^{0}\right)\to\tilde{\beta}_{1}^{\prime}\left(p_{b}^{\prime},\tilde{p}_{b}^{0\prime}\right)= & \tilde{\beta}_{1}\left(p_{b},\frac{\tilde{p}_{b}^{0}}{\xi^{2}}\right),\\
\tilde{\beta}_{2}\left(p_{c},\tilde{p}_{c}^{0}\right)\to\tilde{\beta}_{2}^{\prime}\left(p_{c}^{\prime},\tilde{p}_{c}^{0\prime}\right)= & \tilde{\beta}_{2}\left(p_{c},\tilde{p}_{c}^{0}\right).
\end{align}
It is seen that now $\tilde{\beta}_{2}$ is invariant under rescalings
while $\tilde{\beta}_{1}$ is not. Looking at (\ref{eq:t-horiz})
and (\ref{eq:t-singular}) and the form of $\tilde{\beta}$'s, we
notice that since $\tilde{\beta}_{1}$ becomes important for small
$p_{b}$ (also see Fig. \ref{fig:betas}), it
modifies the behavior near the horizon, while $\beta_{2}$ is important
for the modifications near the singularity because it becomes different
from unity for small $p_{c}$.

\begin{figure}
\noindent \centering{}\includegraphics[scale=.6]{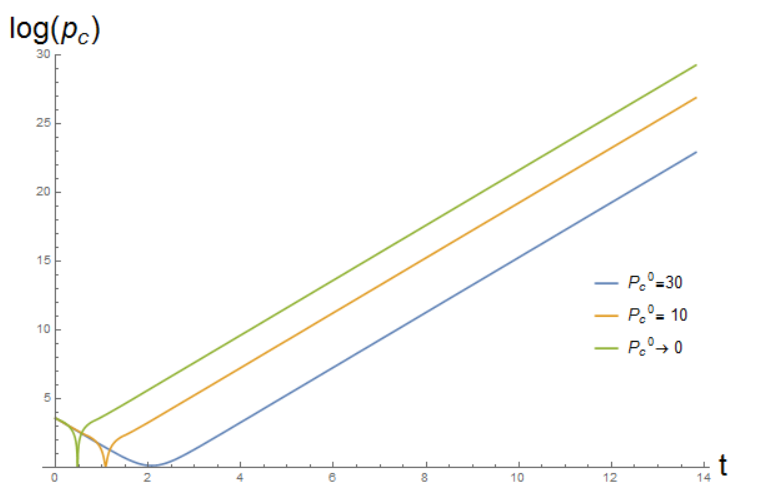}\caption{Effects of the inverse triad corrections on the evolution of $p_{c}(t)$ for distinct
values of $p_{c}^{0}$. The starting point of the curves corresponds to the same black hole mass, while the end of them is associated to the white hole mass as indicated in Eq. \eqref{BHWHmass}. Here we have considered a black hole with mass 2
in Planck units, $L_{0}=10$ and $\gamma=.2375$. These graphs were produced by numerically solving the equations of motion in Python using the Runge–Kutta method of fourth order. The initial and final times are determined by the condition $p_{b}(t_{i})=p_{b}(t_{f})=0$ numerically. The black hole horizon is at $p_{c}(t_{i})$ and the white horizon
at $p_{c}(t_{f})$ where in this plot $t_f\approx 14$. \label{fig efecto}}
\end{figure}

However, for $p_{b}<\tilde{p}_{b}^{0}$ where $\tilde{\beta}_{1}$
becomes important, it will depend on the rescaling. This means that
the classical behavior at the horizon will be affected in this way.
But, if one assumes that one can consider $L_{0}\to\infty$ at the
end, this modification will be avoided (because the region of $p_{b}<\tilde{p}_{b}^{0}$
disappears and we only have $p_{b}>0$ for which $\tilde{\beta}_{1}\to1$
and obviously invariant, see (\ref{eq:beta-1-tilde-1st-Pr})).
Furthermore the quantum corrections at the singularity will not be
modified, since for this limit, $\tilde{\beta}_{2}$ is unaffected.
This limit can be taken for a genuine cosmological model and here
we will assume that it is also valid for black holes. 

The limit $L_0\to\infty$ may also be considered for $\beta$ terms  in Sec V. From Eqs. \eqref{rescal-beta1} and \eqref{rescal-beta2}, it can be seen that this limit eliminates the inverse triad effects of $\beta_2$ which is responsible for corrections at or close to the bounce, since in that limit $\beta_2\to 1$ . But it does not remove the effects of $\beta_1$ which generates  quantum correction at the horizon. Hence, it does not completely remove the dependency of the inverse triad corrections on fiducial quantities.

The first nice consequence of this prescription is that the minimum
value of $p_{c}$ which results in singularity avoidance is now independent
of $L_{0}$,
\begin{equation}
p_{c\text{(min)}}=\gamma GM\sqrt{\Delta}\frac{1}{\sqrt{1-\left(\frac{\ell_{\text{Pl}}}{2GM}\right)^{4}}},\,\,\,\,\,p_{c}>\tilde{p}_{c}^{0}
\end{equation}
while still retaining the factors of modification $\left(1-\left(\frac{\ell_{\text{Pl}}}{2GM}\right)^{4}\right)^{-\frac{1}{2}}$,
which makes the the minimum for the bounce larger than the previous
results. As is seen from the above expression and also from Fig. \ref{masacrit},
this effect is important when $M\approx M_{\textrm{pl}}$. Furthermore, given this value of $p_{c\text{(min)}}$, the curvature scalars are universally bounded.

Using the new tilde variables (\ref{eq:pb0-tilde})-(\ref{eq:beta-2-tilde-1st-Pr})
to compute the expansion and shear, we get similar expression as in
(\ref{eq:theta-orig-N1}) and (\ref{eq:sigma-orig-N1}), but with
replacing $\beta_{1}\to\tilde{\beta}_{1}$, $\beta_{2}\to\tilde{\beta}_{2}$,
$p_{b}^{0}\to\tilde{p}_{b}^{0}$ and $p_{c}^{0}\to\tilde{p}_{c}^{0}$.
As a result, all the terms inside the expressions for $\theta$ and
$\sigma^{2}$, that are proportional to $\tilde{\beta}_{2}$, not
only become independent of $L_{0}$, but also become invariant under
its rescalings. On the other hand, the terms proportional to $\tilde{\beta}_{1}$,  
are not invariant under
its rescalings. However, with our assumption $L_{0}\to\infty$ for
which $\tilde{\beta}_{1}\to 1$, the latter issue is bypassed.
Furthermore, the values of $\theta$ and $\sigma^{2}$ for $N=1$ now
remain finite at both the singularity and the horizon, given that at the horizon,  $\tilde{\beta}_1\to 0$, and $\tilde{\beta}_2$ is finite but nonzero, and at the singularity $\tilde{\beta}_1,\tilde{\beta}_2 \to 0$. In addition, with $L_{0}\to\infty$, they are both invariant.

\begin{figure}
\noindent \begin{centering}
\includegraphics[scale=.6]{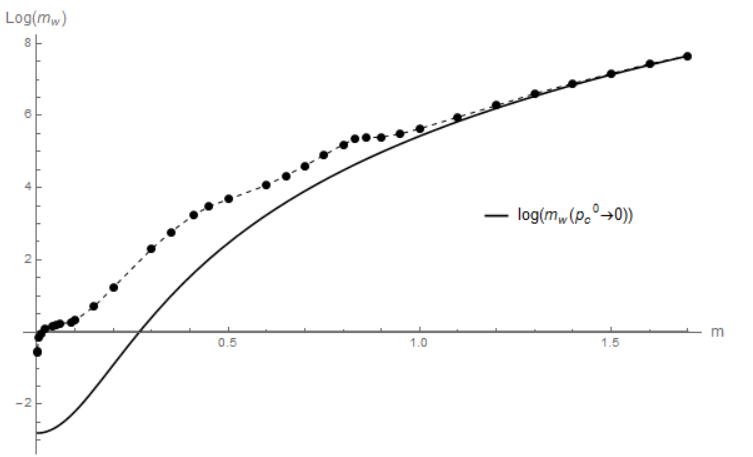}
\par\end{centering}
\caption{Comparing the behavior of $M_{W}\left(M_{B}\right)$ without inverse
triad corrections (solid line) and the case with $\tilde{\beta}_{2}$ corrections.\label{masablanca}}
\end{figure}

Another observation is about the mass of the white hole. The behavior of $p_{c}$
in time, depends on the size of the parameter $\tilde{p}_{c}^{0}$,
and particularly, as can be seen from Fig. \ref{fig efecto}, the
larger the size of $\tilde{p}_{c}^{0}$, the larger the later values
of $p_{c}$. The mass of the black hole and white hole are 
\begin{equation}
M_{B}=M=\frac{\sqrt{p_{c}\left(t_{i}\right)}}{2G},\,\,\,\,\,\,\,\,\,\,\,\,\,\,M_{W}=\frac{\sqrt{p_{c}\left(t_{f}\right)}}{2G},\label{BHWHmass}
\end{equation}
in which $t_{i}$ and $t_{f}$ are the initial and final times when
black hole forms, and then when after bouncing back to the new white
hole with its own Schwarzschild radius, respectively. Here we have
used $p_{c}=r_{\textrm{Schw}}^{2}$ with $r_{\textrm{Schw}}$ being
the Schwarzschild radius of the black and white holes at these two
moments in time. As can be seen from Fig. \ref{fig efecto}, $M_{W}$
then depends on the value of $\tilde{p}_{c}^{0}$ and thus on $\tilde{\beta}_{2}$.
It turns out then that $M_{W}(\beta_{2}\neq1)>M_{W}(\beta_{2}\rightarrow1)$. 
The effect of new correction on the relation between the black hole
and white hole masses can be seen from the graph of $M_{W}\left(M_{B}\right)$
in Fig. \ref{masablanca}.

\subsection{Proposal two: $\delta_{b}\leftrightarrow1/\delta_{c}$ in Thiemann's formula}

The second proposal is to mutually interchange $\delta_{b}\leftrightarrow\frac{1}{\delta_{c}}$, restricted
again to the Thiemann's formula  (\ref{eq:thiemann-trick}). This
leads to the quantities $p_{b}^{0},\,p_{c}^{0},\,\beta_{1}$ and $\beta_{2}$
being replaced by their new versions denoted by $\check{\,}$ as
\begin{align}
\check{p}_{b}^{0}= & \frac{1}{2}\frac{\gamma\ell_{\text{Pl}}^{2}}{\delta_{c}}=\frac{1}{2}\frac{\gamma^{2}\ell_{\text{Pl}}^{4}}{p_{c}^{0}},\\
\check{p}_{c}^{0}= & \frac{\gamma\ell_{\text{Pl}}^{2}}{\delta_{b}}=\frac{1}{2}\frac{\gamma^{2}\ell_{\text{Pl}}^{4}}{p_{b}^{0}},\\
\check{\beta}_{1}\left(p_{b},\check{p}_{b}^{0}\right)= & \frac{\vert p_{b}+\check{p}_{b}^{0}\vert-\vert p_{b}-\check{p}_{b}^{0}\vert}{2\tilde{p}_{b}^{0}},\\
\check{\beta}_{2}\left(p_{c},\check{p}_{c}^{0}\right)= & p_{c}^{\frac{1}{2}}\frac{\left(\sqrt{\vert p_{c}+\check{p}_{c}^{0}\vert}-\sqrt{\vert p_{c}-\check{p}_{c}^{0}\vert}\right)}{\check{p}_{c}^{0}}.
\end{align}
In Fig. \ref{fig:betas12}, the behavior of these new $\check{\beta}_{1},\,\check{\beta}_{2}$ functions is compared to the original functions, and the functions from the first prescription. Interestingly, now both $\check{\beta}_{1},\,\check{\beta}_{2}$ are invariant
under rescalings. Thus, the effective behavior is independent of the auxiliary rescalings both on the horizon and at the singularity, and
thus there is no need for the additional assumption $L_{0}\to\infty$.

\begin{figure}
\noindent \centering{}\includegraphics[scale=.6]{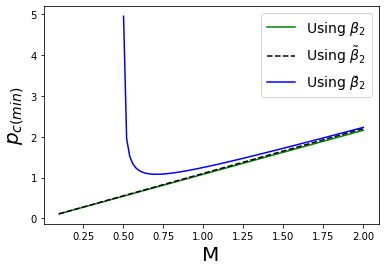}%~~~~~~~~~~~\includegraphics[scale=1.1]{pcminVSpc0-check}
\caption{Dependence of $p_{c\text{(min)}}$ on black hole mass $M$. %On theleft, 
The case without corrections denoted by C-S, is compared to the two prescriptions labeled by their corresponding $\beta_{2}$ functions. Although the lines for the second prescription and no-correction cases look the same, they have  different slopes due to the modification factor $\left(1-\left(\frac{\ell_{\text{Pl}}^{2}}{\Delta}\right)^{2}\right)^{-\frac{1}{2}}$ of the former case. The value of $\Delta$ used here is $\Delta=4\sqrt{3}\pi\gamma\ell_{\text{Pl}}^{2}\approx5.17\ell_{\text{Pl}}^{2}$.  
%On the right, only the second prescription is compared to the case without %corrections, to get a more clear picture.
\label{fig:pcmins-vs}}
\end{figure}

Let us see the effect of these prescription on $p_{c\text{(min)}}$.
Here again, a nice consequence of the prescription is that the minimum value of
$p_{c}$ is independent of $L_{0}$,
\begin{equation}
p_{c\text{(min)}}=\gamma GM\sqrt{\Delta}\frac{1}{\sqrt{1-\left(\frac{\ell_{\text{Pl}}^{2}}{\Delta}\right)^{2}}},\,\,\,\,\,\,\,p_{c}>\check{p}_{c}^{0}.
\end{equation}
Compared to the previous works, this is now modified by a factor $\left(1-\left(\frac{\ell_{\text{Pl}}^{2}}{\Delta}\right)^{2}\right)^{-\frac{1}{2}}$, and unlike proposal one, this factor is now independent of the black hole mass, and instead depends
on the ratio of the Planck area to the minimum area. These modifications
are rather large for any black hole regardless of the mass. In any case, this value of $p_{c\text{(min)}}$, means that in this case too, the curvature scalars are universally bounded. It is worth noting that, in this case, as expected, $p_{c\text{(min)}}\to 0$ if $\Delta\to 0$, which is the case for a theory with a continuous spacetime. Fig, \ref{fig:pcmins-vs} compares the dependence of $p_{c\text{(min)}}$ on black hole mass, $M$, for the two prescriptions as well as the case without the corrections. 

The shear $\sigma^{2}$ and expansion
$\theta$ have similar expression as in (\ref{eq:theta-orig-N1})
and (\ref{eq:sigma-orig-N1}), but with replacing $\beta_{1}\to\check{\beta}_{1}$,
$\beta_{2}\to\check{\beta}_{2}$, $p_{b}^{0}\to\check{p}_{b}^{0}$
and $p_{c}^{0}\to\check{p}_{c}^{0}$. Given that $\check{\beta}_{1}$
and $\check{\beta}_{2}$ are both invariant under rescalings, both
$\theta$ and $\sigma^{2}$ are now fully invariant too, without the need for any further assumptions. Finally they both remain finite
on both horizon and at the classical singularity since at the horizon,  $\check{\beta}_1\to 0$, and $\check{\beta}_2$ is finite but nonzero, and at the singularity $\check{\beta}_1,\check{\beta}_2 \to 0$.

\begin{figure}
\noindent \begin{centering}
\includegraphics[scale=.6]{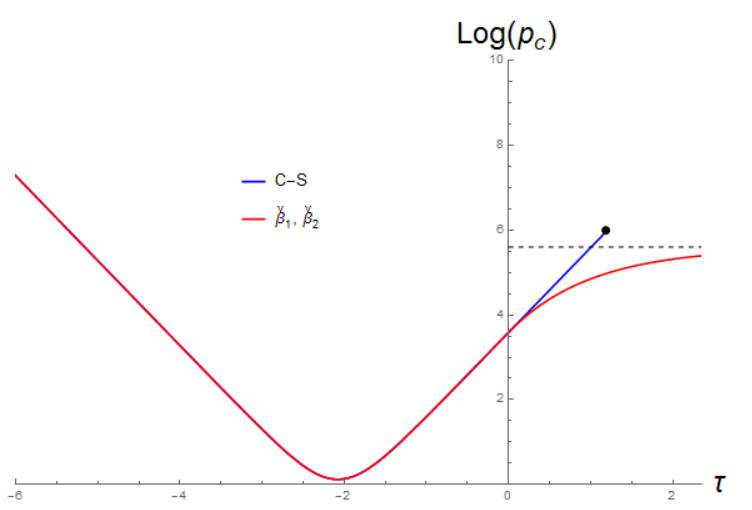}
\par\end{centering}
\caption{The evolution of $p_{c}$ for the case without corrections denoted by C-S, versus the second prescriptions labeled by its corresponding $\check{\beta}_{2}$ functions. Note that the latter case has an asymptote.\label{fig:pc-evolv-check}}
\end{figure}

Finally, the numerical evolution of $p_c$ in time for the case without correction is compared to the second prescription in Fig. \ref{fig:pc-evolv-check}. It is seen that there is a horizontal asymptote corresponding to the case of second prescription, which gives a final value for the $p_c$ which is smaller than the final value without corrections. In the case with no corrections, {\em i.e.}, in \cite{Corichi:2015xia}, the evolution of $p_{c}$ stops at a finite time and the corresponding $p_{c}$  is interpreted as the position of the horizon (the large dot in Fig. \ref{fig:pc-evolv-check}). In the presence of corrections and using the second propsal, however, $p_{c}$ has an asymptotic behavior that depends on $p_{b}^{0}$. Thus this asymptotic value may be interpreted as the horizon, although it is not reached at a finite time.

\section{Discussion\label{sec:Conclusion}}

The quest for a quantum theory of gravity involves looking for physical imprints of the merging of quantum and gravity effects in systems like black holes that classically exhibit  singularities according to general relativity. In the case of Schwarzschild black hole, loop quantum gravity predicts the resolution of the classical singularity as well as a bouncing scenario connecting the black hole to a white hole, a phenomenon also predicted for  some cosmological models, in such a case connecting a contracting with an expanding phase of the Cosmos. Previous research on the loop quantized black hole interior, either purely quantum \cite{Ashtekar:2005qt} or effective \cite{Boehmer:2008fz,Modesto:2005zm,Corichi:2015xia,Cortez:2017alh,Ashtekar:2018lag,Ashtekar:2018cay}, shares this feature. However more insight has been obtained with the latter in that it allowed to investigate the physical independence from auxiliary parameters required in the formulation of the theory  to further improve the description of the black to white hole bounce. Yet, all these works ignore the effects of the inverse triad corrections that enter the Hamiltonian constraint,   basically to simplify the analysis. 

In this work, we have extended and built up over the previous important
works about the effective theory of the interior of the Schwarzschild
modeled as a Kantowski-Sachs spacetime bouncing and with a resolved
singularity. We started by using the proposal for the fiducial cell
parameters in \cite{Corichi:2015xia} to cope with the noncompact topology $\mathbb{R}\times \mathbb{S}^2$ of the model, and used a polymer path integral
approach as systematic way to derive an effective Hamiltonian which automatically incorporates the inverse triad corrections.
The effective Hamiltonian without these corrections resembles exactly the one in previous
works. Although the inclusion of these corrections sheds more light on the
physical nature of the problem, it raises some well-known issues in the analysis if one makes the usual choice of the parameters entering the model $\delta_b, \delta_c$ in both curvature sector as well as in the inverse triad sector \cite{Corichi:2015xia}.  Such issues were the main reason they 
have been mostly ignored in previous works. These issues include the
dependence of physical quantities like the expansion and shear on
the rescaling of the fiducial parameter that is used to define the
fiducial cell. Introduction of this cell is necessary to be able to
compute the symplectic structure. In this case, the cell is a three
dimensional cylinder whose height $L_{0}$ is the fiducial parameter,
while its base is identified with a physical parameter, namely, the
classical Schwarzschild radius. Furthermore, although the ``minimum
radius of black hole at bounce'', $p_{c(\textrm{min})}$, that we obtain here is different
 from the previous works, it also depends on the fiducial parameter $L_{0}$ and its rescalings, which thus renders the model physically unacceptable.

We tackle these issues by exploring the source of the dependence on the auxiliary parameters.
It turns out that triad corrections, contained into two functions
$\beta_{1}$ and $\beta_{2}$, are both noninvariant under their rescalings.
The first function $\beta_{1}$ is important for the behavior of the black hole at the
horizon, while the second one, $\beta_{2}$, is important for the
behavior at or near the singularity. Looking more closely into the
form of these correction functions, we see that some sort of interchanging
of the parameters $\delta_{b}$ and $\delta_{c}$, might resolve the issue; recall these 
are used in the representation of the inverse
triads in Thiemann's formula, as well as in computing the curvatures.  Further inspection reveals that if one only interchanges the parameters $\delta_{b}$ and $\delta_{c}$
in Thiemann's formula, one could make $\beta_{1}$
and $\beta_{2}$ invariant, and furthermore, nothing in the computation
of the curvature changes. In other words only the terms
inside $\beta_{1}$ and $\beta_{2}$ are affected, which is the effect we are looking for.

Given this insight, we study two proposals. The first one consists
of an interchange $\delta_{b}\leftrightarrow\delta_{c}$. As a first
result, the minimum ``radius'' at the bounce $p_{c(\textrm{min})}$,
 now becomes independent of the auxiliary parameter and is modified because
of the presence of the new corrections, such that its value is now
larger than the value computed in previous works. This modification
is particularly important when the mass of the black hole is comparable
to the Planck's mass. Furthermore, the function $\beta_{2}$ also becomes invariant. To investigate this case further, we consider  a congruence of geodesic observers and show that the terms in the corresponding expansion $\theta$ and shear $\sigma^{2}$ proportional
to $\beta_{2}$ are now also invariant. However, in this proposal,
$\beta_{1}$ remains noninvariant. But the issues that correspond
to this function not being invariant can be bypassed by assuming that at
the end, one can take the limit $L_{0}\to\infty$ which may be interpreted
as the black hole being eternal. This limit is legitimate in cosmology
but in a black hole setting, we should be more careful about its consequences.
Using this limit, $\theta$ and $\sigma^{2}$ become both completely
invariant under rescalings. This state of affairs can be improved as we next discusse it.

The second proposal amounts to the interchange $\delta_{b}\leftrightarrow1/\delta_{c}$ in the original choice.
It turns out in this case both $\beta_{1}$ and $\beta_{2}$ are invariant
under rescalings without any need for additional assumptions such
as $L_{0}\to\infty$. As for the congruence of geodesic observers, both expansion $\theta$ and shear
$\sigma^{2}$ now become fully (and not partially) invariant. Another
new result with this proposal is that not only the minimum ``radius''
at the bounce $p_{c(\textrm{min})}$ is different, but now it does
not depend on the mass of the black hole, but rather depends on the ratio of
the minimum area to the Planck area, and is universal for all masses.
This, in general, is certainly larger than what we derived from the previous
proposal, particularly in cases where the mass of the black hole
is much larger than the Planck's mass.  \textit{}
 
Based on our results, one can see that due to the presence of  $\beta_1$, the classical behavior near the horizon will change due to quantum effects.  Some of the other studies of black hole interior in loop quantum gravity have also found similar quantum effects at the horizon \cite{Bohmer:2007wi} \cite{Chiou:2008nm}, although without considering inverse triad corrections. Furthermore, in  \cite{Haggard:2014rza}, and in some of the other scenarios such as the firewall proposal \cite{Almheiri:2012rt}, these quantum effects near the horizon take a significant role. In our case, the correction term $\beta_1$ is responsible for any such quantum effects at the horizon. However, to fully understand and analyze the role and the impact of this correction on the modifications at the horizon we need to extend the model to describe both the interior and the exterior of the black hole. This is the subject of our future investigation.

Finally, we notice that the issue of mismatch between the masses of
initial black hole $M_{B}$, and the final white hole $M_{W}$, worsens
by introducing these new corrections such that $M_{W}(\beta_{2}\neq1)>M_{W}(\beta_{2}\rightarrow1)$.
As mentioned before, this is due to the anisotropic nature of the
Kantowski-Sachs model.

It is worth noting some of the recent works about the same subject. In \cite{Ashtekar:2018cay} the authors expand the interior description to the full spacetime of the Schwarzschild black hole. However, a number of issues that arise in the asymptotic region in this model. The authors also do not consider the inverse triad corrections, although they put forward interesting results regarding the mass of large black holes (such as the relation between black and white holes). They also consider a different presecritions for $\delta_{i}$ parameters, so the effects of the inverse triad corrections will be different and need to be further studied. Also, based on our results, the inverse triad effects are reflected in the relation between the black and white hole masses even for small black holes. In this regards, in \cite{Ashtekar:2018cay}, one may need to include such inverse triad corrections to be able to correctly analyze small black holes. One can also analyze the black hole exterior in their model and check if there exist any significant quantum gravitational effect happening there.

In both \cite{Gambini:2020nsf} and \cite{Kelly:2020uwj},  the authors do not start from a homogeneous space-time, so they do not resort to fiduciary parameters to define the dynamics, and the inverse triad corrections would not have a ``rescaling'' problems. Both works simplify the constraints before quantization by renaming variables and partial gauge fixing, so it is not clear how to implement the inverse triad corrections. Even so, \cite{Gambini:2020nsf} seeks to represent non-polynomial operators similar to the inverse triads, so one possibility is to introduce the inverse triad corrections by adapting the Thiemann trick. On the other hand, \cite{Kelly:2020uwj} simply does not consider these corrections arguing that in some cosmological cases they do not represent significant effects, but it is clear that this is not necessarily valid for other models.
Given what we mentioned, it could be interesting to study the structure of inverse triad corrections in both of these works, since they would not depend on fiducial parameters, and to study possible effects in regions of interest. This will help expand what we know about these corrections so far.

Our results open the door to incorporating further quantum inverse triad corrections in works mentioned above and similar ones, as well as other quantum gravitational systems, hence providing a systematic method to deal with the issues raised by such corrections, which were mostly ignored in previous works. 

\acknowledgments{

The authors would like to acknowledge the support from the CONACyT
Grant No. 237351. S.R. would like to thank the CONACyT SNI support 59344. J.C.R. acknowledges the support of the grant from UAM.  H.A.M.T.  acknowledges the kind hospitality of the Physics department of the ESFM, IPN, during his sabbatical year

}
\bibliography{main}

\end{document}